\documentclass{article}
\usepackage{iclr2027_conference,times}

\usepackage{amsmath,amssymb,amsfonts,mathtools}

\usepackage[utf8]{inputenc}
\usepackage[T1]{fontenc}
\usepackage{url}
\usepackage{booktabs}
\usepackage{multirow}
\usepackage{nicefrac}
\usepackage{microtype}
\usepackage{xcolor}
\usepackage{graphicx}
\usepackage{subcaption}
\usepackage{wrapfig}     
\usepackage{pifont}      
\usepackage{amssymb}     
\usepackage{enumitem}    
\usepackage{etoolbox} 

\usepackage{hyperref}
\usepackage[capitalize,noabbrev]{cleveref}

\newcommand{\E}{\mathbb{E}}

\newcommand{\hz}{\,\mathrm{Hz}}

\iclrfinalcopy

\title{CaDiS: Causal Diffusion Transformer for\\ Streaming Speech Generation}

\author{
Baotong Tian$^{1,2,\dagger}$,
Ke Chen$^{1}$,
Yunyun Wang$^{1}$,
Haonan Chen$^{1}$,
Zeyu Jin$^{1}$ \\
$^{1}$Adobe Research \quad
$^{2}$University of Rochester
}

\begin{document}
\makeatletter
\newcommand{\rev}[1]{{%
  \color{blue}%
  \apptocmd{\@floatboxreset}{\color{blue}}{}{}%
  \captionsetup{%
    font+={color=blue},%
    labelfont+={color=blue},%
    textfont+={color=blue}%
  }%
  #1%
}}
\makeatother
\maketitle
\fancyhead{}
\fancyhead[L]{Preprint}
\begingroup
\renewcommand{\thefootnote}{\fnsymbol{footnote}}
\footnotetext[2]{Work done while at Adobe.}
\endgroup
\begin{abstract}
Diffusion and flow-matching models have advanced high-fidelity speech generation. However, their bidirectional attention requires generating a full utterance before playback, and iterative sampling adds further delay. Most streaming systems therefore rely on autoregressive generators and use diffusion only for acoustic decoding. In this paper, we ask instead whether a pretrained offline diffusion model can generate speech chunk by chunk in real time. We present \textbf{CaDiS}, an offline-to-streaming framework built on a causal diffusion transformer that decouples the architectural and acceleration challenges into a three-stage recipe: autoregressive diffusion adaptation finetunes the bidirectional model with block-causal attention to generate each chunk from past speech alone; causal consistency distillation reduces sampling to four steps; and distribution matching refinement restores teacher-level quality. We adapt CaDiS to two streaming tasks, text-to-speech (TTS) and voice conversion (VC), with dedicated task-specific efforts: a normalized positional encoding for asynchronously aligned text in TTS, and content-vector distillation plus reference-anchored positional re-encoding to address the voice consistency and artifact challenges in VC. We evaluate CaDiS on objective and subjective metrics across natural human and expressive anime voices. CaDiS preserves offline-teacher quality while streaming: its TTS model attains a $286$~ms first-packet latency and $0.31$ real-time factor on an A100 GPU, while its VC model achieves the best intelligibility and perceptual quality among streaming baselines with no look-ahead. Moreover, CaDiS attains among the best overall streaming efficiency profiles, demonstrating the potential of diffusion speech generators in streaming scenarios\footnote{Audio samples can be found at \href{https://cadis.pages.dev}{https://cadis.pages.dev}.}.

\end{abstract}

\section{Introduction}
\label{sec:intro}

Recent advances in generative modeling have driven remarkable progress in speech generation, producing results of high naturalness, intelligibility, and faithfulness~\citep{du2024cosyvoice2,chen2024f5tts,liu2024seedvc}. Building on this progress, streaming speech generation has shown growing promise for interactive applications such as audio previewing, conversational voice agents, and instant voice design. Unlike offline generation, however, it imposes stricter requirements on latency, playback pacing, and the ability to sustain generation in open-ended conversation.

Diffusion models have emerged as leading high-fidelity speech generators. Different from autoregressive (AR) generators built on discrete acoustic tokens, they model continuous acoustic representations, thereby avoiding the fidelity loss introduced by the vector quantization that code-based AR systems adopt for compression~\citep{meng2025autoregressive}. Together with the iterative refinement, diffusion models yield strong naturalness, speaker similarity, and expressiveness~\citep{le2023voicebox,ju2024naturalspeech3,eskimez2024e2tts,chen2024f5tts}. However, such strengths come at the cost of streamability: the bidirectional architectures typically require the full acoustic context, making incremental generation challenging, and the iterative sampling adds further delay. As a result, current streaming systems remain dominated by AR generators, often with diffusion used for local acoustic refinement~\citep{du2024cosyvoice2,du2025cosyvoice3,guo2025streamflow,choi2025streamflow,
wang2024valle2,anastassiou2024seedtts}. Recent approaches have further demonstrated autoregressive generation on continuous speech representations using diffusion~\citep{liu2024ardit,jia2025ditar,wang2025felle,peng2025vibevoice,liu2026audiocalm}.
This leaves the quality advantages of diffusion underused and raises a central question: \emph{can a pretrained diffusion model itself generate speech chunk by chunk at low latency, without sacrificing its offline quality?}

Adapting an offline diffusion model for streaming poses three challenges: \textbf{causality}, \textbf{sampling efficiency}, and \textbf{long-form stability}. Causality requires generating each speech chunk without future acoustic context. Sampling efficiency requires reducing the number of denoising steps. Long-form stability requires the model to sustain quality over continuous, potentially unbounded streams rather than fixed-length utterances.  These challenges take task-specific forms, because TTS and VC differ in how their conditioning aligns with the output. In TTS, text and speech
are asynchronously aligned, so the model must maintain global text-speech alignment across chunks. In VC, the source content is 
frame-synchronous with the output and arrives continuously, so the model must preserve acoustic continuity and speaker identity throughout an open-ended stream using only bounded past context. A streaming framework must therefore address all three challenges while respecting the conditioning structure of each task.

\begin{figure}[t]
\centering
\includegraphics[width=0.95\linewidth]{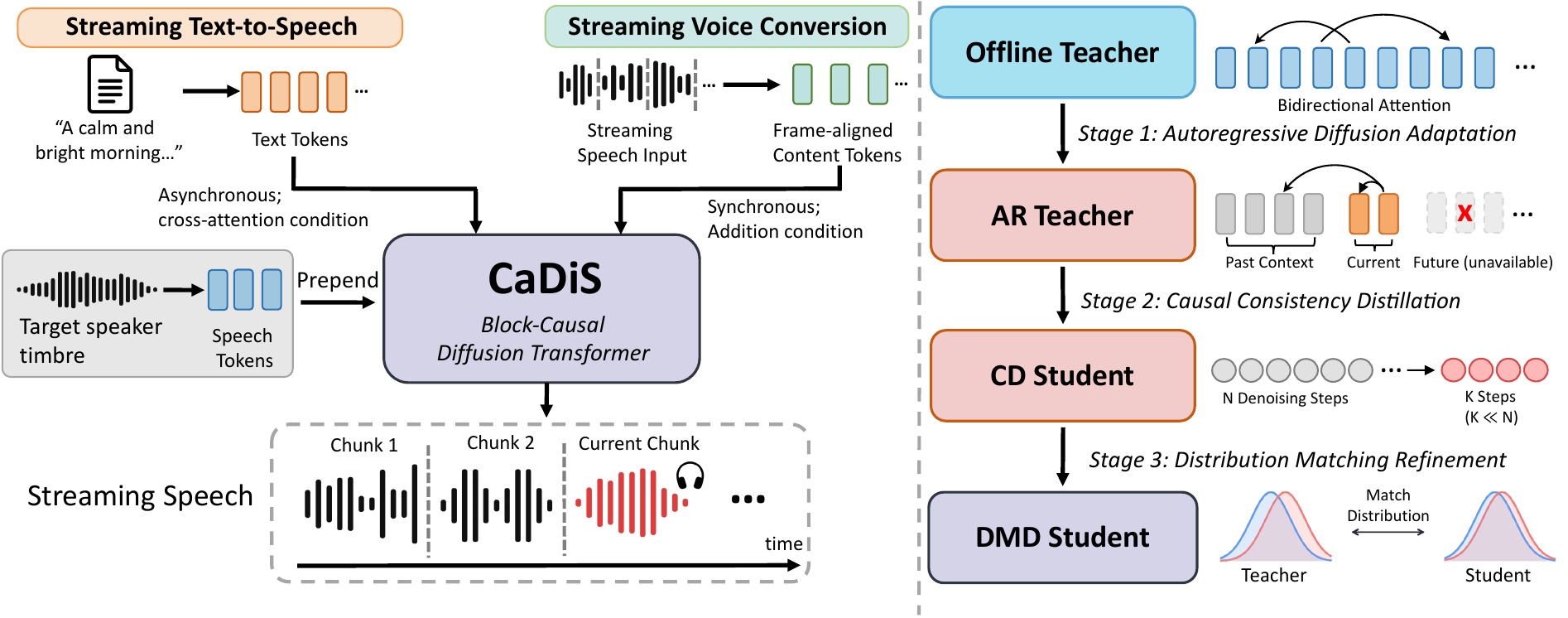}
\caption{The proposed CaDiS architecture and the three-stage recipe. Left: how CaDiS serves streaming TTS and VC tasks. Right: the three-stage recipe to adapt the offline teacher through autoregressive diffusion adaptation (AR Teacher), causal consistency distillation (CD Student) and distribution matching refinement (DMD Student as the final streaming model).}
\label{fig:overview}
\vspace{-5mm}
\end{figure}

We present \textbf{CaDiS}, which adapts a pretrained diffusion transformer (DiT)~\citep{peebles2023dit} speech generator for low-latency streaming while keeping DiT as its generative core (Figure~\ref{fig:overview}). 
CaDiS addresses the streaming-adaptation challenges above through a three-stage recipe. First, autoregressive diffusion adaptation finetunes the pretrained DiT with block-causal attention so that each chunk is generated from past speech alone. Next, causal consistency distillation reduces the denoising process to four steps. Finally, distribution matching refinement restores the distilled streaming model to teacher-level generation quality.
The recipe builds on Causal Forcing++~\citep{zhao2026causalforcingpp}, originally developed for video generation, which we adapt to speech while addressing noticeable and distinct temporal conditioning structures of speech tasks. For TTS, we retain the offline teacher’s normalized positional encoding to accommodate the asynchronous text-speech alignment. For VC, we distill a streamable content-vector encoder~\citep{qian2022contentvec} that produces frame-synchronous content features aligned with the acoustic latents.
And we further adapt reference-anchored positional re-encoding (RAPR)~
\citep{sun2026streamavatar} by pairing a long speaker reference with a short rolling history to sustain speaker identity from bounded context over streams. 

Our contributions are three-fold:
\begin{itemize}[leftmargin=1.4em,itemsep=2pt,topsep=2pt]
\vspace{-1mm}
\item To our knowledge, CaDiS is among the first streamable DiT speech generators operating on continuous audio latents. The three-stage recipe adapts a pretrained offline model for both streaming TTS and VC tasks. 
\item We develop dedicated task-specific designs on CaDiS for streaming TTS and VC tasks: a globally normalized positional encoding for TTS, and reference-anchored positional re-encoding (RAPR) together with content-vector distillation for VC to mitigate the voice consistency and artifact issues. 
\vspace{-1mm}
\item  Extensive objective and subjective evaluations demonstrate that CaDiS produces $48$~kHz audio with competitive overall quality on TTS and the best overall quality on VC among streaming baselines, while achieving a competitive streaming efficiency profile in terms of first-packet latency and real-time factor.
\end{itemize}

\section{Related Work}
\label{sec:related}


\paragraph{Diffusion and flow-matching speech generation.}
Continuous-latent diffusion and flow models have become strong offline speech generators~\citep{le2023voicebox,ju2024naturalspeech3,eskimez2024e2tts,chen2024f5tts}, but their reliance on bidirectional context and iterative sampling hinders streaming.
Recent work explores continuous autoregressive generation: ARDiT combines block-autoregressive diffusion with DMD~\citep{liu2024ardit}, while DiTAR couples a causal LM with a local diffusion transformer~\citep{jia2025ditar}.
FELLE, VibeVoice, and AudioCALM further explore token-wise flow matching, diffusion heads, and shared-transformer denoising, respectively~\citep{wang2025felle,peng2025vibevoice,liu2026audiocalm}.
Unlike these approaches, CaDiS focuses on adapting a pretrained bidirectional speech DiT into a causal, few-step streaming generator.
\paragraph{Streaming text-to-speech.}
Most streaming TTS systems are autoregressive or hybrid, using a language model to predict speech tokens followed by a local acoustic decoder~\citep{wang2024valle2,anastassiou2024seedtts,du2024cosyvoice2,du2025cosyvoice3}. Recent work makes this decoding stage streamable via chunk or block-wise flow-matching~\citep{du2024cosyvoice2,guo2025streamflow,choi2025streamflow}, while Dragon-FM combines autoregressive generation across codec-token chunks with flow matching within each chunk~\citep{liu2025dragonfm}. CaDiS shares this block-autoregressive perspective, but starts from a bidirectional continuous-latent diffusion generator and directly causalizes it into a few-step streaming model, 
generating speech from full-sentence text rather than jointly consuming incremental text and emitting audio \citep{dang2024livespeech2,sheng2025syncspeech,torgashov2025voxtream}.

\paragraph{Streaming voice conversion.}
Existing streaming VC systems typically rely on causal content encoders, streaming-aware training, or language-model conditioning outside the diffusion family~\citep{yang2024streamvc,ning2023dualvc,li2024dualvc3,wang2024streamvoice}. Diffusion and flow models have only recently entered this setting: Seed-VC offers a streaming mode but does not turn an offline generator into a block-causal, few-step primary streamer, while X-VC and MeanVC-2 cut latency using short future buffers~\citep{liu2024seedvc,zheng2026xvc,ma2026meanvc2}. Unlike TTS, VC has no sentence boundaries, so frame-level content alignment and stable speaker identity must hold over long rollouts. Whether zero-look-ahead VC and streaming TTS can share a single causal generator design thus remains open.

\paragraph{Causal and few-step diffusion.}
Diffusion Forcing and rolling diffusion introduce temporally structured noise for sequential generation~\citep{chen2024diffusionforcing,ruhe2024rolling}. CausVid and Self-Forcing distill bidirectional video models into block-causal generators~\citep{yin2024causvid,huang2025selfforcing}. Causal Forcing++ most directly motivates our three-stage recipe~\citep{zhao2026causalforcingpp}. In speech, consistency and distribution-matching methods target few-step synthesis rather than streaming~\citep{ye2023comospeech,ye2024flashspeech,li2024dmospeech}. We ask whether causal distillation can adapt a strong offline speech DiT to both tasks. CaDiS addresses these concerns via task-specific designs.
\section{Method}
\label{sec:method}

Figure~\ref{fig:overview} illustrates the architecture of CaDiS and the three-stage recipe used to train it. 
CaDiS adapts a pretrained offline DiT speech generator into a streaming one. 
The backbone is shared across TTS and VC tasks, but we train a separate task-specific instance for each, distilled from its corresponding offline teacher. We train CaDiS through three stages: autoregressive diffusion adaptation, causal consistency distillation, and distribution matching refinement. 
This section is organized as follows: the streaming generation problem formulation in Section~\ref{sec:formulation}; the architecture of the offline speech generator in Section~\ref{sec:backbone}; the details of the three-stage causalization procedure in Section~\ref{sec:causalize}; and the task-specific designs for TTS and VC in Sections~\ref{sec:tts}--\ref{sec:inference}.

\subsection{Problem formulation}
\label{sec:formulation}

Let $\mathbf z=(\mathbf z^{(1)},\ldots,\mathbf z^{(K)})$ denote a clean audio latent sequence partitioned into $K$ chunks of $L$ frames, where $\mathbf z^{(k)}$ is the $k$-th chunk and $\mathbf z^{(<k)}$ denotes all preceding chunks. Let $\mathbf r$ be a target-speaker reference and $\mathbf u$ the task-dependent content input. While the offline teacher $p_{\mathrm{off}}$ models the full latent sequence bidirectionally, a streaming model generates it chunk by chunk as:
\begin{align}
    \resizebox{0.70\textwidth}{!}{$
    q_\theta(\mathbf z \mid \mathbf r,\mathbf u)
=
\prod_{k=1}^{K}
q_\theta\left(
\mathbf z^{(k)}
\mid
\mathbf r,\mathcal H_k,\mathbf u_k
\right),
\qquad
\mathcal H_k=\mathcal C_k(\mathbf z^{(<k)}).$}
\label{eq:factorization}
\end{align}
Here, $\mathcal C_k$ selects the previously generated audio context, and $\mathbf u_k$ denotes the content available to chunk $k$. For TTS, $\mathbf u_k$ is the text of the current sentence, which is available in full before generation begins. For VC, it is the source-content representation aligned with chunk $k$. Our goal is to learn these causal chunk conditionals such that their streaming rollout approximates the offline distribution.

\subsection{Offline diffusion speech generator}
\label{sec:backbone}

Our pretrained offline generator is a stack of DiT blocks~\citep{peebles2023dit} operating on $40$~Hz continuous latents encoded from $48$~kHz mono audio by a frozen DAC-VAE trained with target-KL regularization~\citep{seetharaman2026targetkl}. 
Both TTS and VC tasks share two conditioning signals: a \emph{speaker reference} that specifies the target timbre, and a \emph{content reference} that specifies what is spoken. For both tasks, the speaker reference is a speech clip that is encoded into audio latents and prepended to the output latent sequence (excluded from the training loss). The content reference is task-specific. For TTS, the content reference is text transcript, which is encoded by an mT5-large text encoder \citep{xue2021mt5} into text tokens for cross-attention condition of the DiT blocks. For VC, the content reference is another speech utterance, which is encoded by a bidirectional 
$50$~Hz ContentVec encoder~\citep{qian2022contentvec}, resampled to $40$~Hz to match the output latent frame rate, and added frame-wise to the output latent sequence. Note that the content reference differs fundamentally between the two tasks: for TTS it is asynchronous text specifying only the linguistic content, whereas for VC it is a frame-aligned representation and the VC model must follow the content frame-by-frame.

The model is trained with a conditional flow-matching objective~\citep{lipman2023flowmatching} that regresses the velocity field of a variance-preserving and shifted-cosine probability path.
Training uses classifier-free guidance~\citep{ho2022cfg}, and the generated latents are decoded to audio by DAC-VAE.

\subsection{From bidirectional to causal diffusion}
\label{sec:causalize}

The offline model is trained with bidirectional attention over the full acoustic context. Simply applying a causal mask alters the conditional distribution to Equation~\ref{eq:factorization}, and rolling the masked model out over many chunks compounds this mismatch into audible drift. To address this, we first adapt the block-causal architecture and then accelerate its sampling by reducing the number of function evaluations (NFEs), with each stage below serving a distinct role.

\paragraph{Stage 1: Autoregressive Diffusion Adaptation.}
Stage 1 addresses the architectural gap to meet the streaming requirement. We replace the bidirectional mask with a block-causal mask and finetune the DiT to predict each chunk from \emph{clean preceding chunks} (i.e., teacher-forcing). Each teacher-forcing example
is arranged as $[\,\mathbf r\mid\mathbf z\mid\mathbf z_t\,]$, where $\mathbf r$ is the speaker
reference, and $\mathbf z$ and $\mathbf z_t$ are clean and noisy speech sequences of the \emph{same} target
utterance. For chunk $k$, we use the noise schedule $(\alpha_t,\sigma_t)$ inherited from the offline teacher, to construct the noisy input and
velocity target with noise $\epsilon \sim \mathcal N(\mathbf 0,\mathbf I)$, and optimize the objective:

\vspace{-0.85cm}
\begin{align}
\label{eq:ar_teacher}
\resizebox{0.65\textwidth}{!}{$
\left.
\mathcal L_{\mathrm{AR}}(\phi)=
\mathbb E_{k,t,\boldsymbol\epsilon}\!\left[
\left\|\mathbf v_\phi(\mathbf z_t^{(k)},t\mid\mathbf c_k)-\mathbf v_t^{*(k)}\right\|_{\mathbf M_k}^2
\right]
\;\middle|\;
\begin{aligned}
\mathbf z_t^{(k)}&=\alpha_t\mathbf z^{(k)}+\sigma_t\boldsymbol\epsilon,\\
\mathbf v_t^{*(k)}&=\alpha_t\boldsymbol\epsilon-\sigma_t\mathbf z^{(k)},\\
\mathbf c_k&\triangleq(\mathbf r,\mathcal H_k,\mathbf u_k).
\end{aligned}
\right.$}
\end{align}
Where $t$ denotes diffusion time throughout, $\phi$ denotes the parameters of this block-causal multi-step model, $\mathbf M_k$ and $\mathbf M$ are binary validity masks for chunk $k$ and the full utterance, respectively. We use $\|\cdot\|_{\mathbf M_k}$ and $\operatorname{mean}_{\mathbf M}$ to denote averages over valid and non-padded elements. $\mathbf c_k$ collects all conditioning visible to chunk $k$. We use a custom block-causal attention mask that preserves the clean preceding chunks required for teacher forcing while preventing future-context leakage. 
By the masking design, all chunks can be trained in parallel while ensuring that each prediction uses only the preceding context.
The resulting model $\mathbf v_\phi$ learns teacher-forced causal chunk conditionals and has the same sampling cost as the offline teacher during inference.

\paragraph{Stage 2: Causal Consistency Distillation.}
Stage 2 closes the sampling-cost gap. We adapt
Causal Forcing++'s online causal consistency objective
\citep{song2023consistency,zhao2026causalforcingpp} to distill the frozen causal teacher $\mathbf v_\phi$ into a four-step causal student $\mathbf v_\theta$. Initialized from the Stage~1 teacher, we define the corresponding clean-latent prediction of the student as $F_\theta(\mathbf z_t,t\mid\mathbf c_k)=\alpha_t\mathbf z_t-\sigma_t\mathbf v_\theta(\mathbf z_t,t\mid\mathbf c_k)$. For adjacent noise levels $t_i>t_{i-1}$, the teacher first denoises the sample by one deterministic ODE step
$\Psi_\phi^{t_i\rightarrow t_{i-1}}$. We then match the student's clean prediction at $t_i$ to that of
its exponential-moving-average target $\bar\theta$ at $t_{i-1}$:
\begin{align}
\label{eq:causal_cd}
\resizebox{0.8\textwidth}{!}{$
\left.
\mathcal L_{\mathrm{CCD}}(\theta)=
\mathbb E\!\left[\left\|
F_\theta(\mathbf z^{(k)}_{t_i},t_i\mid\mathbf c_k)-
\operatorname{sg}\!\left[F_{\bar\theta}(\widetilde{\mathbf z}^{(k)}_{t_{i-1}},t_{i-1}\mid\mathbf c_k)\right]
\right\|_{\mathbf M_k}^2\right]
\;\middle|\;
\widetilde{\mathbf z}^{(k)}_{t_{i-1}}
=\Psi_\phi^{t_i\rightarrow t_{i-1}}\!\left(
\mathbf z^{(k)}_{t_i}\mid\mathbf c_k\right).
\right.$}
\end{align}
Where $\bar\theta$ is an EMA copy of the student parameters $\theta$, and $\operatorname{sg}$ denotes stop-gradient. Both branches share the same teacher-forced causal context, while only the current chunk is noisy. The resulting four-step student initializes Stage~3.

\paragraph{Stage 3: Distribution Matching Refinement.}
Stage 3 recovers the quality lost during few-step distillation. We perform DMD~\citep{yin2024dmd,yin2024dmd2} on student-forced rollouts. The causal student thus learns under its own generated context while matching the distribution of the offline teacher.
Let $G_\theta$ denote the block-causal model parameterized by the Stage~2 student $\mathbf v_\theta$, and let
$\widetilde{\mathbf z}_0=G_\theta(\boldsymbol\epsilon\mid\mathbf r,\mathbf u)$
be an utterance generated chunk by chunk.
We re-noise the student rollout to $\widetilde{\mathbf z}_t$. Given the same noisy sample, the frozen offline teacher provides the real estimate $\widehat{\mathbf z}_{0,\mathrm r}$, while a fake-score model initialized from the offline teacher and updated on student samples provides the fake estimate $\widehat{\mathbf z}_{0,\mathrm f}$. We optimize the following loss:
\begin{align}
\label{eq:dmd}
\resizebox{0.80\textwidth}{!}{$
\mathbf g_{\mathrm{DMD}}
=\frac{\widehat{\mathbf z}_{0,\mathrm f}(\widetilde{\mathbf z}_t,t)
-\widehat{\mathbf z}_{0,\mathrm r}(\widetilde{\mathbf z}_t,t)}
{\operatorname{mean}_{\mathbf M}\bigl|\widetilde{\mathbf z}_0-\widehat{\mathbf z}_{0,\mathrm r}\bigr|},
\qquad
\mathcal L_{\mathrm{DMD}}(\theta)=\frac{1}{2}\E\!\left[
\left\|\widetilde{\mathbf z}_0-
\operatorname{sg}\!\left[\widetilde{\mathbf z}_0-\mathbf g_{\mathrm{DMD}}\right]\right\|_{\mathbf M}^2
\right].$}
\end{align}
After each chunk is generated, a detached copy is added to the causal cache, so subsequent chunks condition on student-generated rather than ground-truth history, without back-propagating through previously generated chunks. 
This matches the history seen during streaming inference while preserving four-step generation.

\subsection{Streaming TTS}
\label{sec:tts}
\paragraph{Text Condition and Global Alignment.}
For streaming TTS, the content condition is the raw text of \emph{the current sentence}, injected into each acoustic chunk through cross-attention. We generate and stream the speech incrementally. Because the offline TTS model learns a global monotonic relationship between textual and acoustic progress, resetting the positional information during the generation process disrupts this alignment prior learned during pretraining. We therefore retain the normalized absolute positional encoding of our offline teacher when adapting to streaming. An external duration predictor estimates the total utterance length once from the sentence text, and each acoustic frame is then assigned its fractional position within this predicted length. This preserves sentence-level progress across streaming chunks and allows the block-causal model to maintain text-speech alignment without access to future acoustic context.

\paragraph{Reference-based Duration Estimation.}
We estimate the target utterance length using a reference-based speaking-rate heuristic rather than a separately trained duration predictor. Specifically, we compute the speaker's average number of acoustic latent frames per IPA token from the reference speech and its transcript. We then apply this rate to the IPA-token count of each target sentence, with additional adjustments for punctuation and non-speech events. During training, we use the ground-truth utterance length for position normalization. At inference, the length is estimated once per sentence before generation, determining the total number of acoustic latent frames to be generated by the DiT model.
\subsection{Streaming voice conversion}
\label{sec:vc}

\paragraph{Temporally Synchronous Content Condition.}
To enable frame-synchronous conditioning, the content condition and the output latents must share the same frame rate. In our offline VC system, 
we interpolate the $50$~Hz bidirectional ContentVec representation to $40$~Hz to match the output latents. When adapting to a streaming model, we apply two optimizations: (1) we initialize a native $40$~Hz ContentVec encoder for content condition, by relocating the convolutional strides of the pretrained $50$~Hz encoder so that the same weights produce $40$~Hz features; and (2) we make ContentVec block-causal by removing the look-ahead from both the convolutional and transformer layers. The resulting causal ContentVec encoder is trained to match the $50$~Hz bidirectional ContentVec teacher (interpolated at $40$~Hz output). As a result, the streaming VC model natively produces content and latent streams at the same rate, without a separate alignment step.

\paragraph{Long-form Streaming with RAPR.}
Unlike TTS, streaming VC has no sentence boundary to reset generation. The model must therefore preserve local prosodic continuity and target-speaker identity over an open-ended stream while keeping positional indices bounded.
We adapt reference-anchored positional re-encoding (RAPR)~\citep{sun2026streamavatar} with a long speaker reference and a short rolling history. Let $\mathbf r$ be the reference, $w$ the history-window length, and $n_k$ the number of frames generated before chunk $k$. At chunk $k$, we keep the most recent
$\min(n_k,w)$ frames as $\mathcal R_k^w$ and form the attention cache
$\mathcal K_k^w=[\mathbf r\mid\mathcal R_k^w]$. Keys are cached before positional rotation, so positions can be reassigned as the window slides. The position map $\pi_k$ preserves the reference anchors and re-indexes the retained history into the slots right after the reference. Once the stream exceeds $w$ frames, the oldest frame is dropped, bounding the cache size and positional span at $|\mathbf r|+w$. Section~\ref{sec:longform} studies the effect of $w$ length on long-form generation quality and speaker timbre consistency.

\vspace{-0.25cm}
\subsection{Streaming inference}
\label{sec:inference}

For streaming TTS, continuous generation can be re-anchored because the text input contains natural linguistic boundaries. We therefore apply a sentence-aware inference mechanism that discards the accumulated DiT positional state and preceding latents at each sentence boundary, then generate the next sentence independently from the speaker reference. A short overlap in the decoder output smooths the acoustic transition between consecutive sentences. This also reduces error accumulation of the target speaker timbre during long-form generation.

For streaming VC, we apply RAPR streaming to run generation as a continuous loop. For each incoming chunk, we first compute the $40$~Hz ContentVec features, then assemble the RAPR context, generate the next chunk, decode it to a waveform and update the causal cache. Our main configuration uses $400$~ms chunks, a $1$~s history window, 4 NFEs, and no look-ahead. Together with the chunk-causal ContentVec encoder, this forms a fully causal VC pipeline.
\vspace{-0.2cm}
\section{Experiments}
\label{sec:exp}

\vspace{-0.1cm}
\subsection{Experimental setup}
\label{sec:setup}

\paragraph{Training and Evaluation Setup. } 

CaDiS is trained on a large corpus of internal and publicly licensed speech data. The streaming TTS models are trained on $311$k hours of speech data over three stages. The streaming VC models are trained on $179$k hours in Stage 1 and $110$k hours in each of Stage 2 and 3. We verify that no training data overlaps with the evaluation benchmarks described.

We assess zero-shot streaming TTS and cross-speaker streaming VC tasks on two benchmarks across normal human voices and expressive anime voices:
\begin{itemize}[leftmargin=1.4em,itemsep=2pt,topsep=2pt]
\item Seed-TTS-eval~\citep{anastassiou2024seedtts}: the English \texttt{test-en} split of 1088 CommonVoice  utterances for TTS and a non-parallel VC split of 1086 source-target utterance pairs. 
\item Anim400K~\citep{cai2024anim400k}: we select 1000 English anime clips to create 1000 TTS and VC pairs. 
\end{itemize}
To support long-form evaluation in Section~\ref{sec:longform}, we additionally construct $20$ VC sources by concatenating consecutive utterances from the LibriSpeech \texttt{test-clean} set, yielding an average duration of $124$\,s. Each source is paired with a Seed-TTS-eval reference voice. 
We mainly report English evaluations. Mandarin and other results are provided along with the test-set details in appendices.

\paragraph{Model Setup.}
Both tasks use the same backbone with a separate model trained for each task. 
Following the three-stage training recipe, the final streaming model is the \emph{Stage~3 DMD student}. 
The $0.8$B-parameter offline bidirectional teacher is trained for $500$k steps. The subsequent causal stages require $50$k steps for the AR teacher on $32$ A100 GPUs, $30$k steps for the consistency student on $32$ GPUs, and only $1000$ steps for the DMD student on $64$ GPUs. Additional data and hyperparameter details are provided in Appendix~\ref{app:impl}.

\paragraph{Baselines.}
We generate outputs for all baselines from their released weights. For TTS, we compare against offline baselines F5-TTS, IndexTTS-2.5, and Fish-Speech S1-mini~\citep{chen2024f5tts,deng2025indextts2,liao2024fishspeech}; and 
the streaming baselines CosyVoice 3, Qwen3-TTS, VoXtream2, and Faster-IndexTTS-2~\citep{du2025cosyvoice3, hu2026qwen3tts, torgashov2025voxtream, du2026fasterindextts2}.
For VC, we compare against the offline baseline Vevo~\citep{zhang2025vevo} and streaming baselines Seed-VC, X-VC, and MeanVC-2~\citep{liu2024seedvc, zheng2026xvc, ma2026meanvc2}. Stage-wise ablation results are provided in Section~\ref{sec:ablation}, Appendices~\ref{app:ensim} and~\ref{app:zh}.

\paragraph{Objective Metrics.} We adopt word error rate (WER)~\citep{radford2023whisper}, speaker similarity via WeSpeaker~\citep{wang2023wespeaker}, and UTMOS~\citep{saeki2022utmos}. SIM$_\text{WeSp}$ uses the SimAM-ResNet100 checkpoint\footnote{https://github.com/wenet-e2e/wespeaker/blob/master/docs/pretrained.md}.

\paragraph{Subjective Metrics.}
We conduct subjective listening tests to evaluate models across both tasks. We use a gender-balanced subset of 200 clips per task and benchmark to let participants rate on a 1-5 scale on three questions: naturalness (NMOS), perceptual quality (QMOS), and speaker similarity (SMOS). For VC, we additionally ask participants to assess source-rhythm (\textsc{Rhy}) and source-expression (\textsc{Exp}) preservation between the source and generated utterances.
We collected around 400 submissions per TTS task (total 1200) by 200 participants and 300 submissions per VC task (total 900) by 100 participants. 
Full protocols and results are provided in Appendix~\ref{app:subjective}. 

\paragraph{Streaming Metrics.}
To assess the streaming efficiency, we report first-packet latency (FPL) and real-time factor (RTF).
The first-packet latency (FPL) of VC additionally includes the wall-clock compute required to produce the first output chunk, whereas TTS FPL is simply the wall-clock time to the first output packet. The RTF is computed as $\text{wall time}/\text{speech duration}$, where RTF$<$1 indicates real-time throughput. All timings are measured on an idle A100 GPU. Additional VC ablations on streaming assessments are reported in Appendices~\ref{app:vcablate}--\ref{app:steps}.

\subsection{Streaming TTS Results}
\label{sec:tts_results}

Table~\ref{tab:tts_en} reports streaming TTS results on the Seed-TTS-eval English benchmark. Due to the page limit, the corresponding results on Anim400K are deferred to Table~\ref{tab:anim400k_tts} in Appendix~\ref{app:ood}.

Relative to our offline teacher, CaDiS retains comparable performance on both objective and subjective metrics using only 4 denoising steps. WER improves slightly from $1.75$ to $1.69$, while SIM$_{\text{WeSp}}$ decreases from $0.816$ to $0.785$ and UTMOS from $4.29$ to $4.19$. The gap on subjective tests is smaller still, with a modest gain in naturalness and a slight reduction in speaker similarity. Overall, CaDiS largely preserves the perceptual quality of its offline counterpart.

\begin{table}[t]
\centering\small
\setlength{\tabcolsep}{4pt}
\caption{Streaming TTS on Seed-TTS-eval English across objective and subjective metrics.}
\label{tab:tts_en}
\resizebox{0.85\textwidth}{!}{%
\begin{tabular}{lcccc|ccc}
\toprule
& & \multicolumn{3}{c|}{Objective} & \multicolumn{3}{c}{Subjective MOS} \\
\cmidrule(lr){3-5}\cmidrule(lr){6-8}
System & Stream & WER\,$\downarrow$ & SIM$_{\text{WeSp}}$\,$\uparrow$ & UTMOS\,$\uparrow$ & NMOS\,$\uparrow$ & QMOS\,$\uparrow$ & SMOS\,$\uparrow$ \\
\midrule
\multicolumn{8}{l}{\textit{Offline / Segment-level references}}\\
\midrule
\textbf{Ours} --- offline & \ding{55} & 1.75 & 0.816 & 4.29 & 3.90 & 4.17 & 3.85 \\
F5-TTS                                   & \ding{55} & 1.91 & 0.815 & 4.15 & 3.78 & 4.01 & 3.79 \\
IndexTTS-2.5                             & \ding{55} & 2.58 & 0.791 & 3.98 & 3.62 & 3.80 & 3.68 \\
S1-mini                    & \ding{55}  & 1.72 & 0.697 & 4.37 & 3.65 & 4.15 & 3.34\\
\midrule
\multicolumn{8}{l}{\textit{Streaming systems}}\\
\midrule
Qwen3-TTS                                & \checkmark & \textbf{1.57} & 0.769 & \textbf{4.43} & \underline{3.91} & \underline{4.09} & 3.69 \\
VoXtream2                                & \checkmark & 1.98 & \underline{0.803} & 3.98 & 3.62 & 3.92 & 3.78 \\
CosyVoice3                               & \checkmark & 2.48 & 0.802 & \underline{4.25} & 3.51 & 4.00 & 3.68 \\
Faster-IndexTTS-2                        & \checkmark & 2.41 & \textbf{0.813} & 3.95 & 3.75 & 3.82 & \textbf{3.80} \\
\textbf{CaDiS (ours)}                 & \checkmark & \underline{1.69} & 0.785 & 4.19 & \textbf{3.99} & \textbf{4.17} & \underline{3.79} \\
\bottomrule
\end{tabular}}
\vspace{-0.5cm}
\end{table}
Among the streaming baselines, CaDiS is particularly strong under human evaluation, achieving the best naturalness and audio quality scores, and the second-best speaker similarity score. On objective metrics, it achieves the second-lowest WER and the third-highest UTMOS score. While its speaker similarity trails the strongest baselines, it remains within a competitive range. Overall, these results indicate that CaDiS offers a favorable balance of intelligibility and human-rated perceptual quality. \


\subsection{Streaming Voice Conversion Results}
\label{sec:vc_results}

Table~\ref{tab:vc_en} reports the streaming VC results. Relative to our offline teacher, CaDiS improves both WER and speaker similarity (SIM$_\text{WeSp}$) but drops in UTMOS. The subjective results follow the same trend. 
However, among the streaming baselines, CaDiS achieves the far lowest WER, the highest UTMOS, and superior scores to nearly all subjective measures except SMOS. The main trade-off but also noteworthy observation is speaker similarity: X-VC achieves higher similarity ($0.748$ vs.\ $0.670$), though the two systems are close in SMOS ($3.16$ vs.\ $3.10$). Source preservation evaluations further reveal a complementary advantage of CaDiS. It achieves the highest rhythm ($4.00$) and expression ($3.82$) scores among the streaming baselines, closely matching its offline teacher. 

We further evaluate VC performance on Anim400K in Table~\ref{tab:anim400k_vc}, whose expressive anime voices are entirely out-of-domain and span a broader range of timbres and styles. Among the streaming baselines, CaDiS again achieves the lowest WER, the highest QMOS, and second-best scores of all other metrics. CaDiS performs similarly to X-VC but yields substantially better perceptual quality in both UTMOS and QMOS. 

Above all, while CaDiS does not maximize speaker similarity, it delivers the best intelligibility and perceptual quality, as well as preserves the best timing and expressive characteristics of source speech across in-domain and out-of-domain voices. 




\begin{table}[t]
\centering\small
\setlength{\tabcolsep}{4pt}
\caption{Streaming VC on Seed-TTS-eval English across objective and subjective metrics.} 

\label{tab:vc_en}
\resizebox{0.82\textwidth}{!}{%
\begin{tabular}{lccc|ccccc}
\toprule
& \multicolumn{3}{c|}{Objective} & \multicolumn{5}{c}{Subjective MOS} \\
\cmidrule(lr){2-4}\cmidrule(lr){5-9}
System & WER$\downarrow$ & SIM$_{\mathrm{WeSp}}\uparrow$ & UTMOS$\uparrow$ & NMOS$\uparrow$ & QMOS$\uparrow$ & SMOS$\uparrow$ & \textsc{Rhy}$\uparrow$ & \textsc{Exp}$\uparrow$ \\
\midrule
\multicolumn{9}{l}{\textit{Offline references}} \\
\midrule
\textbf{Ours} --- offline & 3.11 & 0.653 & 4.11 & 3.92 & 3.87 & 3.01 & 3.95 & 3.81 \\
Vevo & 3.09 & 0.756 & 3.96 & 3.96 & 3.95 & 3.67 & 3.95 & 3.71 \\
\midrule
\multicolumn{9}{l}{\textit{Streaming systems}} \\
\midrule
Seed-VC & 5.85 & \underline{0.675} & 3.67 & 3.32 & 2.99 & 2.90 & 3.65 & 3.43 \\
X-VC & \underline{3.41} & \textbf{0.748} & 3.45 & 3.40 & 2.84 & \textbf{3.16} & \underline{3.85} & \underline{3.64} \\
MeanVC-2 & 6.20 & 0.632 & \underline{3.79} & \underline{3.58} & \underline{3.26} & 2.37 & 3.73 & 3.52 \\
\textbf{CaDiS} & \textbf{3.09} & 0.670 & \textbf{3.90} & \textbf{3.79} & \textbf{3.73} & \underline{3.10} & \textbf{4.00} & \textbf{3.82} \\
\bottomrule
\end{tabular}}
\vspace{-4pt}
\end{table}


\begin{table}[t]

\centering\small

\setlength{\tabcolsep}{4pt}

\caption{Streaming VC on Anim400K (out-of-domain) across objective and subjective metrics.}
\vspace{-2mm}
\label{tab:anim400k_vc}

\resizebox{0.82\textwidth}{!}{%

\begin{tabular}{lcccc|ccc}

\toprule

& & \multicolumn{3}{c|}{Objective} & \multicolumn{3}{c}{Subjective MOS} \\

\cmidrule(lr){3-5}\cmidrule(lr){6-8}

System & Stream & WER\,$\downarrow$ & SIM$_{\text{WeSp}}$\,$\uparrow$ & UTMOS\,$\uparrow$ & NMOS\,$\uparrow$ & QMOS\,$\uparrow$ & SMOS\,$\uparrow$ \\

\midrule

\multicolumn{8}{l}{\textit{Offline references}}\\

\midrule
\textbf{Ours} --- offline & \ding{55} & 2.88 & 0.702 & 4.19 & 3.74 & 3.82 & 3.29 \\

Vevo & \ding{55} & 3.42 & 0.750 & 4.13 & 3.83 & 3.87 & 3.76 \\

\midrule

\multicolumn{8}{l}{\textit{Streaming systems}}\\

\midrule

Seed-VC & \checkmark & 4.18 & 0.693 & \textbf{4.09} & 3.19 & \underline{3.64} & 2.92 \\

X-VC & \checkmark & \underline{3.45} & \textbf{0.741} & 3.84 & \textbf{3.60} & 3.19 & \textbf{3.28} \\

MeanVC-2 & \checkmark & 5.64 & 0.655 & 3.95 & 3.28 & 3.20 & 2.59 \\
\textbf{CaDiS} & \checkmark & \textbf{3.11} & \underline{0.695} & \underline{4.04} & \underline{3.58} & \textbf{3.73} & \underline{3.12} \\
\bottomrule
\end{tabular}}
\vspace{-6pt}
\end{table}

\begin{table}[t]
\centering\small
\caption{First-packet latency (FPL), real-time factor (RTF), and look-ahead (LA) measured on A100.}
\vspace{-2mm}
\label{tab:latency}
\begin{subtable}[b]{0.46\textwidth}
\centering
\setlength{\tabcolsep}{5pt}
\caption{Streaming TTS}
\begin{tabular}{lcc}
\toprule
System & FPL\,(ms)\,$\downarrow$ & RTF\,$\downarrow$ \\
\midrule
Qwen3-TTS & 334 & 0.39 \\
VoXtream2 & \textbf{183} & 0.32 \\
CosyVoice3 & 923 & 0.34 \\
Faster-IndexTTS-2 & 553 & \textbf{0.30} \\
\textbf{CaDiS}& \underline{286} & \underline{0.31} \\
\bottomrule
\end{tabular}
\end{subtable}
\hfill
\begin{subtable}[b]{0.52\textwidth}
\centering
\setlength{\tabcolsep}{5pt}
\caption{Streaming VC}
\begin{tabular}{lccc}
\toprule
System & LA\,(ms)\,$\downarrow$ & FPL\,(ms)\,$\downarrow$ & RTF\,$\downarrow$ \\
\midrule
Seed-VC & \underline{20} & 466 & 0.58 \\
X-VC & 100 & \underline{285} & 0.35 \\
MeanVC-2 & 40 & \textbf{211} & \textbf{0.20} \\
\textbf{CaDiS} & \textbf{0} & 575 & \underline{0.26} \\
\bottomrule
\end{tabular}
\end{subtable}
\vspace{-6pt}
\end{table}

\subsection{Streaming Efficiency}
Table~\ref{tab:latency} further reports the streaming efficiency across both tasks. For TTS, CaDiS achieves $286$~ms first-packet latency and $0.31$ RTF, outperforming Qwen3-TTS and CosyVoice3. VoXtream2 has lower FPL but higher RTF, while Faster-IndexTTS-2 has lower RTF but a substantially higher FPL ($553$~ms). For VC, CaDiS trades a high FPL ($575$~ms) for zero-look-ahead operation while retaining the second-best RTF. MeanVC-2 exhibits stronger efficiency while its generation quality falls far below CaDiS and other baselines, making the efficiency gain a poor trade. X-VC, in contrast, incurs both a higher RTF and a less favorable combination of look-ahead and FPL than CaDiS, and does so while producing lower-quality speech. 

Overall, these results show that CaDiS offers among the best overall streaming efficiency across baselines, and still delivers the best perceptual quality and intelligibility of speech generation.

\subsection{Training Stage decomposition}
\label{sec:ablation}

Table~\ref{tab:ablation} decomposes the contribution of each training stage, and a consistent pattern emerges across both TTS and VC. Stage 1 introduces block causality, opening a modest quality gap to the offline teacher. Stage 2 reduces sampling steps from 32 to 4 at the cost of a further quality drop. Stage~3 DMD then recovers most of the quality loss at no extra inference cost. On TTS, Stage 2 trades WER and UTMOS, which are improved back by Stage 3, with the speaker similarity even improved on VC.  
Overall, the three stages play distinct and effective roles by enabling streaming capability, reducing inference cost, and restoring quality, respectively. They together form an integral pipeline of offline-to-streaming adaptation for a high-quality diffusion speech generator.

\begin{table}[t]
\centering\small
\setlength{\tabcolsep}{6pt}
\caption{The ablation study of three training stages evaluated on Seed-TTS-eval English. }
\vspace{-2mm}
\label{tab:ablation}
\resizebox{\textwidth}{!}{%
\begin{tabular}{llccc|ccc}
\toprule
& & \multicolumn{3}{c|}{TTS} & \multicolumn{3}{c}{VC} \\
Stage & Steps & WER\,$\downarrow$ & SIM$_{\text{WeSp}}$\,$\uparrow$ & UTMOS\,$\uparrow$ & WER\,$\downarrow$ & SIM$_{\text{WeSp}}$\,$\uparrow$ & UTMOS\,$\uparrow$ \\
\midrule
Offline teacher (reference)      & --  & 1.75 & \textbf{0.816} & \textbf{4.29} & 3.11 & 0.653 & \textbf{4.11} \\
Stage~1: AR causal teacher     & 32  & 1.73 & 0.807 & 4.14 & 3.20 & 0.660 & 3.88 \\
\;+Stage~2: consistency (CD)   & 4   & 2.05 & 0.767 & 3.82 & 3.23 & 0.639 & 3.76 \\
\;+Stage~3: DMD (CaDiS)     & 4   & \textbf{1.69} & 0.785 & 4.19 & \textbf{3.09} & \textbf{0.670} & 3.90 \\
\bottomrule
\end{tabular}}
\vspace{-4mm}
\end{table}

\subsection{Long-form streaming}
\label{sec:longform}

\begin{wrapfigure}{r}{0.50\textwidth}
  \centering
  \vspace{-\intextsep}
  \includegraphics[width=\linewidth]{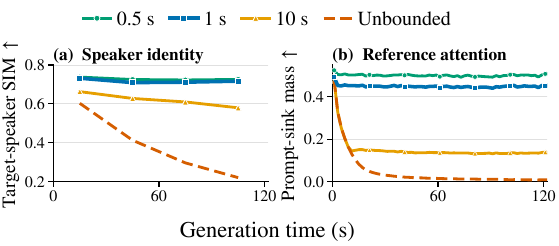}
  \caption{RAPR prevents speaker timbre drift in long-form streaming VC.}
  \label{fig:rapr}
  \vspace{-3mm}
\end{wrapfigure}
\vspace{-2mm}
We study long-form speaker identity control through VC, where the target speaker must be preserved
continuously from a fixed reference. Figure~\ref{fig:rapr}(a) shows the effect of the RAPR history window when streaming long sequence. Without bounding the
generated history, target-speaker similarity heavily degrades over time. A $10$\,s window
reduces but does not eliminate this drift, while $0.5$\,s and $1$\,s windows keep identity
nearly constant throughout the rollout. Thus, limiting generated history is critical for
stable long-form VC. To understand why, we further obtain the decoder attention map assigned to the target prompt sink.
For each generated query position, we sum its normalized attention weights over all keys that belong to the pinned target prompt, giving the fraction of attention the model assigns to the speaker reference. We then average this fraction across query positions, attention heads, decoder layers, and the four sampling steps to obtain one anchor-attention value for each generation chunk. As shown in
Figure~\ref{fig:rapr}(b), short RAPR windows maintain a large, stable attention mass on the
target reference, whereas longer or unbounded history progressively shifts attention away from the target. This trend
closely tracks the speaker drift in Figure~\ref{fig:rapr}(a), indicating that bounded history
preserves identity by sustaining attention to the reference. Overall, RAPR is essential for stable long-form VC. We use a $1$\,s window, which matches the
long-form identity stability of $0.5$\,s while giving the best short-form WER
(Appendix~\ref{app:rapr}).

\vspace{-1.5mm}
\section{Conclusion}
\label{sec:conclusion}
\vspace{-1mm}
Streaming speech generation does not require abandoning high-quality diffusion generators in favor of autoregressive models. We showed that a pretrained bidirectional speech Diffusion Transformer can be adapted into four-step generators through autoregressive diffusion adaptation, causal consistency distillation, and distribution matching refinement. The same backbone design supports two structurally different tasks: streaming TTS, where normalized positional encoding preserves text--speech alignment under block-causal generation, and streaming voice conversion, where frame-synchronous content conditioning and reference-anchored positional re-encoding maintain speaker identity under bounded context. Long-form experiments further show that RAPR limits speaker drift over extended streams by sustaining attention to the speaker reference. Across standard English speech and expressive anime speech benchmarks, the resulting streaming models approach their offline teachers while achieving competitive streaming efficiency in terms of first-packet latency and real-time factor. These results demonstrate the potential of high-quality diffusion speech generators for responsive and interactive streaming applications.



\subsubsection*{Reproducibility statement}
We report the exact evaluation tooling (Whisper-large-v3, Paraformer-zh, WavLM-large-SV, Resemblyzer,
SimAM-ResNet100, UTMOS) and run all scoring through the Seed-TTS-eval scripts so numbers are directly
comparable to prior work; every baseline is generated in-house from released weights under the same
scorer; and Section~\ref{sec:setup} specifies the distillation chain as a verified init-from sequence and
names the GPU behind every latency, with warm-up discarded.
Appendix~\ref{app:impl} details the architecture, chunk/RoPE-cache and KV-cache
configuration, per-task recipes, hyperparameters, and data mixes; Appendix~\ref{app:data} details
test-set construction and the loudness/enhancement fairness protocol.


\section*{AI Use Statement}

Generative AI tools were used to assist with research ideation and execution,
drafting portions of the manuscript, language editing, and improving its
readability. They were not used for literature retrieval or discovery,
generating synthetic datasets, or proving mathematical claims. All
AI-assisted ideas, analyses, and text were reviewed and verified by the
authors, who take full responsibility for the content of the paper.

\bibliography{references}

@string{icassp = "Proc. ICASSP"}

@string{interspeech = "Proc. Interspeech"}

@string{icml = "Proc. ICML"}

@string{iclr = "Proc. ICLR"}

@string{iccv = "Proc. ICCV"}

@string{nips = "Proc. NeurIPS"}

@string{slt = "Proc. SLT"}

@string{acm-multimedia = "Proc. ACM Multimed."}

@string{cvpr = "Proc. CVPR"}

@string{acl = "Proc. ACL"}

@string{naacl-hlt = "Proc. NAACL-HLT"}

@string{ncmmsc = "Proc. NCMMSC"}

@misc{anastassiou2024seedtts,
  doi = {10.48550/ARXIV.2406.02430},
  author = {Anastassiou, Philip and Chen, Jiawei and Chen, Jitong and Chen, Yuanzhe and Chen, Zhuo and Chen, Ziyi and Cong, Jian and Deng, Lelai and Ding, Chuang and Gao, Lu and Gong, Mingqing and Huang, Peisong and Huang, Qingqing and Huang, Zhiying and Huo, Yuanyuan and Jia, Dongya and Li, Chumin and Li, Feiya and Li, Hui and Li, Jiaxin and Li, Xiaoyang and Li, Xingxing and Liu, Lin and Liu, Shouda and Liu, Sichao and Liu, Xudong and Liu, Yuchen and Liu, Zhengxi and Lu, Lu and Pan, Junjie and Wang, Xin and Wang, Yuping and Wang, Yuxuan and Wei, Zhen and Wu, Jian and Yao, Chao and Yang, Yifeng and Yi, Yuanhao and Zhang, Junteng and Zhang, Qidi and Zhang, Shuo and Zhang, Wenjie and Zhang, Yang and Zhao, Zilin and Zhong, Dejian and Zhuang, Xiaobin},
  title = {{Seed-TTS}: A Family of High-Quality Versatile Speech Generation Models},
  year = {2024},
  note = {arXiv:2406.02430}
}

@inproceedings{cai2024anim400k,
  author = {Cai, Kevin and Liu, Chonghua and Chan, David M.},
  title = {{ANIM-400K}: A Large-Scale Dataset for Automated End-to-End Dubbing of Video},
  booktitle = icassp,
  pages = {11796--11800},
  year = {2024},
  doi = {10.1109/ICASSP48485.2024.10447911},
}

@misc{shoukanlabs2023anispeech,
  author = {{ShoukanLabs}},
  title = {{AniSpeech} Dataset},
  year = {2023},
  howpublished = {Hugging Face dataset},
  url = {https://huggingface.co/datasets/ShoukanLabs/AniSpeech},
  note = {Accessed September 25, 2026}
}

@inproceedings{wang2023wespeaker,
  author = {Wang, Hongji and Liang, Chengdong and Wang, Shuai and Chen, Zhengyang and Zhang, Binbin and Xiang, Xu and Deng, Yanlei and Qian, Yanmin},
  title = {{WeSpeaker}: A Research and Production Oriented Speaker Embedding Learning Toolkit},
  booktitle = icassp,
  pages = {1--5},
  year = {2023},
  doi = {10.1109/ICASSP49357.2023.10096626},
}

@inproceedings{chen2024diffusionforcing,
  author = {Chen, Boyuan and Mart{\'i} Mons{\'o}, Diego and Du, Yilun and Simchowitz, Max and Tedrake, Russ and Sitzmann, Vincent},
  title = {Diffusion Forcing: Next-token Prediction Meets Full-Sequence Diffusion},
  booktitle = nips,
  volume = {37},
  pages = {24081--24125},
  year = {2024},
  doi = {10.52202/079017-0759},
}

@inproceedings{chen2024f5tts,
  author = {Chen, Yushen and Niu, Zhikang and Ma, Ziyang and Deng, Keqi and Wang, Chunhui and Zhao, Jian and Yu, Kai and Chen, Xie},
  title = {{F5-TTS}: A Fairytaler that Fakes Fluent and Faithful Speech with Flow Matching},
  booktitle = acl,
  pages = {6255--6271},
  year = {2025},
  doi = {10.18653/v1/2025.acl-long.313},
}

@misc{deng2025indextts2,
  author = {Li, Yunpei and Zhou, Xun and Wang, Jinchao and Wang, Lu and Wu, Yong and Zhou, Siyi and Zhou, Yiquan and Shu, Jingchen},
  title = {{IndexTTS 2.5} Technical Report},
  year = {2026},
  note = {arXiv:2601.03888},
  doi = {10.48550/arXiv.2601.03888},
}

@misc{du2024cosyvoice2,
  doi = {10.48550/ARXIV.2412.10117},
  author = {Du, Zhihao and Wang, Yuxuan and Chen, Qian and Shi, Xian and Lv, Xiang and Zhao, Tianyu and Gao, Zhifu and Yang, Yexin and Gao, Changfeng and Wang, Hui and Yu, Fan and Liu, Huadai and Sheng, Zhengyan and Gu, Yue and Deng, Chong and Wang, Wen and Zhang, Shiliang and Yan, Zhijie and Zhou, Jingren},
  title = {{CosyVoice 2}: Scalable Streaming Speech Synthesis with Large Language Models},
  year = {2024},
  note = {arXiv:2412.10117}
}

@inproceedings{eskimez2024e2tts,
  author = {Eskimez, Sefik Emre and Wang, Xiaofei and Thakker, Manthan and Li, Canrun and Tsai, Chung-Hsien and Xiao, Zhen and Yang, Hemin and Zhu, Zirun and Tang, Min and Tan, Xu and Liu, Yanqing and Zhao, Sheng and Kanda, Naoyuki},
  title = {{E2 TTS}: Embarrassingly Easy Fully Non-Autoregressive Zero-Shot {TTS}},
  booktitle = slt,
  pages = {682--689},
  year = {2024},
  doi = {10.1109/SLT61566.2024.10832320},
}

@misc{ho2022cfg,
  doi = {10.48550/ARXIV.2207.12598},
  author = {Ho, Jonathan and Salimans, Tim},
  title = {Classifier-Free Diffusion Guidance},
  year = {2022},
  note = {arXiv:2207.12598}
}

@inproceedings{huang2025selfforcing,
  author = {Huang, Xun and Li, Zhengqi and He, Guande and Zhou, Mingyuan and Shechtman, Eli},
  title = {Self Forcing: Bridging the Train-Test Gap in Autoregressive Video Diffusion},
  booktitle = nips,
  volume = {38},
  pages = {185411--185436},
  year = {2025},
  doi = {10.52202/085713-5576},
}

@misc{zhao2026causalforcingpp,
  doi = {10.48550/ARXIV.2605.15141},
  author = {Zhao, Min and Zhu, Hongzhou and Zheng, Kaiwen and Zhou, Zihan and Yan, Bokai and Li, Xinyuan and Yang, Xiao and Li, Chongxuan and Zhu, Jun},
  title = {Causal Forcing++: Scalable Few-Step Autoregressive Diffusion Distillation for Real-Time Interactive Video Generation},
  year = {2026},
  note = {arXiv:2605.15141}
}

@inproceedings{sun2026streamavatar,
  author = {Sun, Zhiyao and Peng, Ziqiao and Ma, Yifeng and Chen, Yi and Zhou, Zhengguang and Zhou, Zixiang and Zhang, Guozhen and Zhang, Youliang and Zhou, Yuan and Lu, Qinglin and Liu, Yong-Jin},
  title = {{StreamAvatar}: Streaming Diffusion Models for Real-Time Interactive Human Avatars},
  booktitle = cvpr,
  pages = {10887--10897},
  year = {2026},
}

@inproceedings{ju2024naturalspeech3,
  author = {Ju, Zeqian and Wang, Yuancheng and Shen, Kai and Tan, Xu and Xin, Detai and Yang, Dongchao and Liu, Eric and Leng, Yichong and Song, Kaitao and Tang, Siliang and Wu, Zhizheng and Qin, Tao and Li, Xiangyang and Ye, Wei and Zhang, Shikun and Bian, Jiang and He, Lei and Li, Jinyu and Zhao, Sheng},
  title = {{NaturalSpeech 3}: Zero-Shot Speech Synthesis with Factorized Codec and Diffusion Models},
  booktitle = icml,
  volume = {235},
  pages = {22605--22623},
  year = {2024},
  publisher = {PMLR},
}

@inproceedings{le2023voicebox,
  author = {Le, Matthew and Vyas, Apoorv and Shi, Bowen and Karrer, Brian and Sari, Leda and Moritz, Rashel and Williamson, Mary and Manohar, Vimal and Adi, Yossi and Mahadeokar, Jay and Hsu, Wei-Ning},
  title = {Voicebox: Text-Guided Multilingual Universal Speech Generation at Scale},
  booktitle = nips,
  volume = {36},
  pages = {14005--14034},
  year = {2023},
  doi = {10.52202/075280-0618},
}

@inproceedings{li2024dualvc3,
  author = {Ning, Ziqian and Wang, Shuai and Zhu, Pengcheng and Wang, Zhichao and Yao, Jixun and Xie, Lei and Bi, Mengxiao},
  title = {{DualVC 3}: Leveraging Language Model Generated Pseudo Context for End-to-end Low Latency Streaming Voice Conversion},
  booktitle = interspeech,
  pages = {197--201},
  year = {2024},
  doi = {10.21437/Interspeech.2024-1857},
}

@inproceedings{lipman2023flowmatching,
  author = {Lipman, Yaron and Chen, Ricky T. Q. and Ben-Hamu, Heli and Nickel, Maximilian and Le, Matt},
  title = {Flow Matching for Generative Modeling},
  booktitle = iclr,
  year = {2023},
}

@misc{liu2024seedvc,
  doi = {10.48550/ARXIV.2411.09943},
  author = {Liu, Songting},
  title = {Zero-shot Voice Conversion with Diffusion Transformers},
  year = {2024},
  note = {arXiv:2411.09943}
}

@inproceedings{zhang2025vevo,
  author = {Zhang, Xueyao and Zhang, Xiaohui and Peng, Kainan and Tang, Zhenyu and Manohar, Vimal and Liu, Yingru and Hwang, Jeff and Li, Dangna and Wang, Yuhao and Chan, Julian and Huang, Yuan and Wu, Zhizheng and Ma, Mingbo},
  title = {Vevo: Controllable Zero-Shot Voice Imitation with Self-Supervised Disentanglement},
  booktitle = iclr,
  year = {2025},
}

@misc{liao2024fishspeech,
  doi = {10.48550/ARXIV.2411.01156},
  author = {Liao, Shijia and Wang, Yuxuan and Li, Tianyu and Cheng, Yifan and Zhang, Ruoyi and Zhou, Rongzhi and Xing, Yijin},
  title = {{Fish-Speech}: Leveraging Large Language Models for Advanced Multilingual Text-to-Speech Synthesis},
  year = {2024},
  note = {arXiv:2411.01156}
}

@inproceedings{ning2023dualvc,
  author = {Ning, Ziqian and Jiang, Yuepeng and Zhu, Pengcheng and Yao, Jixun and Wang, Shuai and Xie, Lei and Bi, Mengxiao},
  title = {{DualVC}: Dual-mode Voice Conversion using Intra-model Knowledge Distillation and Hybrid Predictive Coding},
  booktitle = interspeech,
  pages = {2063--2067},
  year = {2023},
  doi = {10.21437/Interspeech.2023-1157},
}

@inproceedings{peebles2023dit,
  author = {Peebles, William and Xie, Saining},
  title = {Scalable Diffusion Models with Transformers},
  booktitle = iccv,
  pages = {4195--4205},
  year = {2023},
  doi = {10.1109/ICCV51070.2023.00387},
}

@inproceedings{qian2022contentvec,
  author = {Qian, Kaizhi and Zhang, Yang and Gao, Heting and Ni, Junrui and Lai, Cheng-I and Cox, David and Hasegawa-Johnson, Mark and Chang, Shiyu},
  title = {{ContentVec}: An Improved Self-Supervised Speech Representation by Disentangling Speakers},
  booktitle = icml,
  volume = {162},
  pages = {18003--18017},
  year = {2022},
  publisher = {PMLR},
}

@inproceedings{radford2023whisper,
  author = {Radford, Alec and Kim, Jong Wook and Xu, Tao and Brockman, Greg and McLeavey, Christine and Sutskever, Ilya},
  title = {Robust Speech Recognition via Large-Scale Weak Supervision},
  booktitle = icml,
  volume = {202},
  pages = {28492--28518},
  year = {2023},
  publisher = {PMLR},
}

@inproceedings{saeki2022utmos,
  author = {Saeki, Takaaki and Xin, Detai and Nakata, Wataru and Koriyama, Tomoki and Takamichi, Shinnosuke and Saruwatari, Hiroshi},
  title = {{UTMOS}: {UTokyo-SaruLab} System for {VoiceMOS} Challenge 2022},
  booktitle = interspeech,
  pages = {4521--4525},
  year = {2022},
  doi = {10.21437/Interspeech.2022-439},
}

@inproceedings{seetharaman2026targetkl,
  author = {Seetharaman, Prem and Kumar, Rithesh},
  title = {Taming Audio {VAE}s via Target-{KL} Regularization},
  booktitle = icassp,
  pages = {15487--15491},
  year = {2026},
  doi = {10.1109/ICASSP55912.2026.11460662},
}

@inproceedings{song2023consistency,
  author = {Song, Yang and Dhariwal, Prafulla and Chen, Mark and Sutskever, Ilya},
  title = {Consistency Models},
  booktitle = icml,
  volume = {202},
  pages = {32211--32252},
  year = {2023},
  publisher = {PMLR},
}

@inproceedings{wang2024streamvoice,
  author = {Wang, Zhichao and Chen, Yuanzhe and Wang, Xinsheng and Xie, Lei and Wang, Yuping},
  title = {{StreamVoice}: Streamable Context-Aware Language Modeling for Real-time Zero-Shot Voice Conversion},
  booktitle = acl,
  pages = {7328--7338},
  year = {2024},
  doi = {10.18653/v1/2024.acl-long.396},
}

@misc{wang2024valle2,
  doi = {10.48550/ARXIV.2406.05370},
  author = {Chen, Sanyuan and Liu, Shujie and Zhou, Long and Liu, Yanqing and Tan, Xu and Li, Jinyu and Zhao, Sheng and Qian, Yao and Wei, Furu},
  title = {{VALL-E 2}: Neural Codec Language Models are Human Parity Zero-Shot Text to Speech Synthesizers},
  year = {2024},
  note = {arXiv:2406.05370}
}

@inproceedings{yang2024streamvc,
  author = {Yang, Yang and Kartynnik, Yury and Li, Yunpeng and Tang, Jiuqiang and Li, Xing and Sung, George and Grundmann, Matthias},
  title = {{StreamVC}: Real-Time Low-Latency Voice Conversion},
  booktitle = icassp,
  pages = {11016--11020},
  year = {2024},
  doi = {10.1109/ICASSP48485.2024.10446863},
}

@inproceedings{yin2024causvid,
  author = {Yin, Tianwei and Zhang, Qiang and Zhang, Richard and Freeman, William T. and Durand, Fr{\'e}do and Shechtman, Eli and Huang, Xun},
  title = {From Slow Bidirectional to Fast Autoregressive Video Diffusion Models},
  booktitle = cvpr,
  pages = {22963--22974},
  year = {2025}
}

@inproceedings{yin2024dmd,
  author = {Yin, Tianwei and Gharbi, Micha{\"e}l and Zhang, Richard and Shechtman, Eli and Durand, Fr{\'e}do and Freeman, William T. and Park, Taesung},
  title = {One-Step Diffusion with Distribution Matching Distillation},
  booktitle = cvpr,
  pages = {6613--6623},
  year = {2024}
}

@inproceedings{yin2024dmd2,
  author = {Yin, Tianwei and Gharbi, Micha{\"e}l and Park, Taesung and Zhang, Richard and Shechtman, Eli and Durand, Fr{\'e}do and Freeman, William T.},
  title = {Improved Distribution Matching Distillation for Fast Image Synthesis},
  booktitle = nips,
  volume = {37},
  pages = {47455--47487},
  year = {2024},
  doi = {10.52202/079017-1505},
}

@misc{du2025cosyvoice3,
  doi = {10.48550/ARXIV.2505.17589},
  author = {Du, Zhihao and Gao, Changfeng and Wang, Yuxuan and Yu, Fan and Zhao, Tianyu and Wang, Hao and Lv, Xiang and Wang, Hui and Ni, Chongjia and Shi, Xian and An, Keyu and Yang, Guanrou and Li, Yabin and Chen, Yanni and Gao, Zhifu and Chen, Qian and Gu, Yue and Chen, Mengzhe and Chen, Yafeng and Zhang, Shiliang and Wang, Wen and Ye, Jieping},
  title = {{CosyVoice 3}: Towards In-the-wild Speech Generation via Scaling-up and Post-training},
  year = {2025},
  note = {arXiv:2505.17589}
}

@misc{hu2026qwen3tts,
  doi = {10.48550/ARXIV.2601.15621},
  author = {Hu, Hangrui and Zhu, Xinfa and He, Ting and Guo, Dake and Zhang, Bin and Wang, Xiong and Guo, Zhifang and Jiang, Ziyue and Hao, Hongkun and Guo, Zishan and Zhang, Xinyu and Zhang, Pei and Yang, Baosong and Xu, Jin and Zhou, Jingren and Lin, Junyang},
  title = {{Qwen3-TTS} Technical Report},
  year = {2026},
  note = {arXiv:2601.15621}
}

@misc{du2026fasterindextts2,
  doi = {10.48550/ARXIV.2607.21042},
  author = {Du, Muyang and Yu, Shuang and Lai, Junjie},
  title = {{Faster IndexTTS-2}: Accelerating and Streaming Autoregressive Zero-Shot Text-to-Speech Synthesis on {GPU}s},
  year = {2026},
  note = {arXiv:2607.21042}
}

@misc{torgashov2025voxtream,
  author = {Torgashov, Nikita and Henter, Gustav Eje and Skantze, Gabriel},
  title = {{VoXtream2}: Full-stream {TTS} with Dynamic Speaking Rate Control},
  year = {2026},
  note = {arXiv:2603.13518},
  doi = {10.48550/arXiv.2603.13518},
}

@misc{dang2024livespeech2,
  doi = {10.48550/ARXIV.2410.00767},
  author = {Dang, Trung and Aponte, David and Tran, Dung and Chen, Tianyi and Koishida, Kazuhito},
  title = {Zero-Shot Text-to-Speech from Continuous Text Streams},
  year = {2024},
  note = {arXiv:2410.00767}
}

@inproceedings{sheng2025syncspeech,
  author = {Sheng, Zhengyan and Du, Zhihao and Zhang, Shiliang and Yan, Zhijie and Chen, Liping},
  title = {{SyncSpeech}: Efficient and Low-Latency Text-to-Speech Based on Temporal Masked Transformer},
  booktitle = icassp,
  pages = {16682--16686},
  year = {2026},
  doi = {10.1109/ICASSP55912.2026.11460607},
}

@misc{zheng2026xvc,
  doi = {10.48550/ARXIV.2604.12456},
  author = {Zheng, Qixi and Zhao, Yuxiang and Wang, Tianrui and Chen, Wenxi and Xu, Kele and Li, Yikang and Cheng, Qinyuan and Qiu, Xipeng and Yu, Kai and Chen, Xie},
  title = {{X-VC}: Zero-shot Streaming Voice Conversion in Codec Space},
  year = {2026},
  note = {arXiv:2604.12456}
}

@inproceedings{ma2026meanvc2,
  author = {Ma, Guobin and Xia, Yuxuan and Jiang, Yuepeng and Guo, Dake and Xie, Hanke and Hu, Jingbin and Wang, Yanbo and Xie, Lei and Zhu, Pengcheng},
  title = {{MeanVC 2}: Robust Low-Latency Streaming Zero-Shot Voice Conversion},
  booktitle = interspeech,
  pages = {4721--4725},
  year = {2026},
  doi = {10.21437/Interspeech.2026-1961},
}

@inproceedings{ruhe2024rolling,
  author = {Ruhe, David and Heek, Jonathan and Salimans, Tim and Hoogeboom, Emiel},
  title = {Rolling Diffusion Models},
  booktitle = icml,
  volume = {235},
  pages = {42818--42835},
  year = {2024},
  publisher = {PMLR},
}

@misc{liu2024ardit,
  doi = {10.48550/ARXIV.2406.05551},
  author = {Liu, Zhijun and Wang, Shuai and Inoue, Sho and Bai, Qibing and Li, Haizhou},
  title = {Autoregressive Diffusion Transformer for Text-to-Speech Synthesis},
  year = {2024},
  note = {arXiv:2406.05551}
}

@inproceedings{jia2025ditar,
  author = {Jia, Dongya and Chen, Zhuo and Chen, Jiawei and Du, Chenpeng and Wu, Jian and Cong, Jian and Zhuang, Xiaobin and Li, Chumin and Wei, Zhen and Wang, Yuping and Wang, Yuxuan},
  title = {{DiTAR}: Diffusion Transformer Autoregressive Modeling for Speech Generation},
  booktitle = icml,
  volume = {267},
  pages = {27255--27270},
  year = {2025},
  publisher = {PMLR},
}

@inproceedings{ye2023comospeech,
  author = {Ye, Zhen and Xue, Wei and Tan, Xu and Chen, Jie and Liu, Qifeng and Guo, Yike},
  title = {{CoMoSpeech}: One-Step Speech and Singing Voice Synthesis via Consistency Model},
  booktitle = acm-multimedia,
  pages = {1831--1839},
  year = {2023},
  doi = {10.1145/3581783.3612061},
}

@inproceedings{ye2024flashspeech,
  author = {Ye, Zhen and Ju, Zeqian and Liu, Haohe and Tan, Xu and Chen, Jianyi and Lu, Yiwen and Sun, Peiwen and Pan, Jiahao and Bian, Weizhen and He, Shulin and Xue, Wei and Liu, Qifeng and Guo, Yike},
  title = {{FlashSpeech}: Efficient Zero-Shot Speech Synthesis},
  booktitle = acm-multimedia,
  pages = {6998--7007},
  year = {2024},
  doi = {10.1145/3664647.3681044},
}

@inproceedings{li2024dmospeech,
  author = {Li, Yinghao Aaron and Kumar, Rithesh and Jin, Zeyu},
  title = {{DMOSpeech}: Direct Metric Optimization via Distilled Diffusion Model in Zero-Shot Speech Synthesis},
  booktitle = icml,
  volume = {267},
  pages = {35186--35208},
  year = {2025},
  publisher = {PMLR},
}

@inproceedings{guo2025streamflow,
  author = {Guo, Dake and Yao, Jixun and Ma, Lihan and Wang, He and Xie, Lei},
  title = {{StreamFlow}: Streaming Flow Matching with Block-Wise Guided Attention Mask for Speech Token Decoding},
  booktitle = ncmmsc,
  series = {Communications in Computer and Information Science},
  pages = {75--86},
  year = {2026},
  publisher = {Springer Nature Singapore},
  doi = {10.1007/978-981-95-5382-2_7},
}

@inproceedings{choi2025streamflow,
  author = {Choi, Ha-Yeong and Lee, Sang-Hoon},
  title = {{StreamFlow}: Streaming Audio Generation from Discrete Tokens via Streaming Flow Matching},
  booktitle = nips,
  volume = {38},
  pages = {5243--5271},
  year = {2025},
  doi = {10.52202/085713-0165},
}

@misc{liu2025dragonfm,
  doi = {10.48550/ARXIV.2507.22746},
  author = {Liu, Yanqing and Xue, Ruiqing and Zhang, Chong and Liu, Yufei and Wang, Gang and Li, Bohan and Qian, Yao and He, Lei and Liu, Shujie and Zhao, Sheng},
  title = {Next Tokens Denoising for Speech Synthesis},
  year = {2025},
  note = {arXiv:2507.22746}
}

@inproceedings{meng2025autoregressive,
  author = {Meng, Lingwei and Zhou, Long and Liu, Shujie and Chen, Sanyuan and Han, Bing and Hu, Shujie and Liu, Yanqing and Li, Jinyu and Zhao, Sheng and Wu, Xixin and Meng, Helen M. and Wei, Furu},
  title = {Autoregressive Speech Synthesis without Vector Quantization},
  booktitle = acl,
  pages = {1287--1300},
  year = {2025},
  doi = {10.18653/v1/2025.acl-long.65},
}

@inproceedings{xue2021mt5,
  title     = {m{T5}: A Massively Multilingual Pre-trained Text-to-Text Transformer},
  author    = {Xue, Linting and Constant, Noah and Roberts, Adam and Kale, Mihir and Al-Rfou, Rami and Siddhant, Aditya and Barua, Aditya and Raffel, Colin},
  booktitle = naacl-hlt,
  pages     = {483--498},
  year      = {2021}
}

@inproceedings{wang2025felle,
  author = {Wang, Hui and Liu, Shujie and Meng, Lingwei and Li, Jinyu and Yang, Yifan and Zhao, Shiwan and Sun, Haiyang and Liu, Yanqing and Sun, Haoqin and Zhou, Jiaming and Lu, Yan and Qin, Yong},
  title = {{FELLE}: Autoregressive Speech Synthesis with Token-Wise Coarse-to-Fine Flow Matching},
  booktitle = acm-multimedia,
  year = {2025},
  doi = {10.1145/3746027.3755494},
}

@inproceedings{peng2025vibevoice,
  author = {Peng, Zhiliang and Yu, Jianwei and Wang, Wenhui and Chang, Yaoyao and Sun, Yutao and Dong, Li and Zhu, Yi and Xu, Weijiang and Bao, Hangbo and Wang, Zehua and Huang, Shaohan and Xia, Yan and Wei, Furu},
  title = {{VibeVoice}: Expressive Podcast Generation with Next-Token Diffusion},
  booktitle = iclr,
  year = {2026},
}

@inproceedings{liu2026audiocalm,
  author = {Liu, Huadai and Luo, Kaicheng and Wang, Wen and Chen, Qian and Ma, Bin and Li, Xiangang and Xue, Wei},
  title = {{AudioCALM}: Continuous Autoregressive Language Modeling for Universal Audio Generation},
  booktitle = nips,
  year = {2026},
}
\bibliographystyle{iclr2027_conference}

\newpage
\appendix
\section{Test-set construction and fairness protocol}
\label{app:data}

All short-form results use the public Seed-TTS-eval benchmark~\citep{anastassiou2024seedtts}, which ships a native zero-shot VC task---its
released \texttt{non\_para\_reconstruct} split (a source utterance $+$ a distinct non-parallel speaker as
timbre reference), used verbatim---scored with the same
WER/CER/SIM tooling as its TTS task, so TTS and VC are directly comparable. WER uses Whisper-large-v3
and CER uses Paraformer-zh (FunASR), both via the Seed-TTS-eval scripts verbatim; SIM$_{\text{Res}}$
is Resemblyzer cosine, SIM$_{\text{WavLM}}$ is WavLM-large-SV cosine, and SIM$_{\text{WeSp}}$ is SimAM-ResNet100/VoxBlink2 cosine. Every system converts toward, and is scored against, the
\emph{same} reference voice pre-processed by an enhancement front end. For Seed-TTS-eval the source/content
audio is left unmodified; the anime out-of-domain pools (Appendix~\ref{app:ood}) also enhance the source,
so both source and target are enhanced there. Streaming and offline audio are loudness-normalized before scoring, and all objective
metrics used here are approximately level-invariant. Long-form TTS uses $100$ English passages
($\sim$48\,s) and Mandarin sentence-concatenations ($\sim$145\,s). For the English set, we stitch
disjoint, already-punctuated Seed-TTS-eval sentences into the $100$ passages. Long-form VC uses
LibriSpeech-Long sources ($\sim$123\,s, English) and AISHELL-3 single-speaker sources ($\sim$122\,s,
Mandarin). For English VC, we stitch consecutive utterances read within the same LibriSpeech
\texttt{test-clean} chapter into $20$ sources averaging $\sim$124\,s, then pair each source with a
Seed-TTS-eval reference voice so that the long- and short-form sets have the same target-speaker
distribution.

Our primary anime benchmark is \textbf{Anim400K}~\citep{cai2024anim400k}, approximately $425$K aligned animated video segments
in Japanese and English collected from TV anime episodes; we use only its English audio and apply an
audio-quality, single-speaker-purity, and caption-faithfulness gate to select $1000$ clips, from which we construct $1000$ TTS pairs and $1000$ VC pairs. We additionally include
\textbf{AniSpeech}~\citep{shoukanlabs2023anispeech} ($23{,}656$ clips from $265$ voices) as a second
anime benchmark for breadth and likewise construct $1000$ TTS pairs and $1000$ VC pairs.
\begin{table}[htbp]
\centering\small
\setlength{\tabcolsep}{5pt}
\caption{\textbf{Short-form zero-shot TTS, Seed-TTS-eval English --- full table} (enhanced targets).
Companion to main-text Table~\ref{tab:tts_en} with the Resemblyzer (SIM$_{\text{Res}}$) and
WavLM-large-SV (SIM$_{\text{WavLM}}$) similarity variants (the main text reports WeSpeaker), plus the AR-teacher and CD rows.}
\label{tab:tts_en_full}
\resizebox{\textwidth}{!}{%
\begin{tabular}{lccccccc}
\toprule
System & Stream & WER\,$\downarrow$ & SIM$_{\text{Res}}$\,$\uparrow$ & SIM$_{\text{WavLM}}$\,$\uparrow$ & UTMOS\,$\uparrow$ & FPL\,(ms)\,$\downarrow$ & RTF\,$\downarrow$ \\
\midrule
\multicolumn{8}{l}{\textit{Offline references}}\\
\midrule
\textbf{Ours} --- offline & \ding{55} & 1.75 & 0.910 & 0.677 & 4.29 & -- & -- \\
Qwen3-TTS (offline)                                & \ding{55} & 1.52 & 0.886 & 0.670 & 4.43 & -- & -- \\
F5-TTS                                   & \ding{55} & 1.91 & 0.897 & 0.657 & 4.15 & -- & -- \\
Faster-IndexTTS-2 (offline)                        & \ding{55} & 2.04 & 0.898 & 0.665 & 4.04 & -- & -- \\
CosyVoice3 (offline)                               & \ding{55} & 1.91 & 0.906 & 0.697 & 4.26 & -- & -- \\
IndexTTS-2.5                             & \ding{55} & 2.58 & 0.904 & 0.607 & 3.98 & -- & -- \\
S1-mini                                   & \ding{55} & 1.72 & 0.861 & 0.579 & 4.37 & -- & -- \\
\midrule
\multicolumn{8}{l}{\textit{Streaming systems}}\\
\midrule

Qwen3-TTS                                & \checkmark & \textbf{1.57} & 0.880 & 0.659 & \textbf{4.43} & 334 & 0.39 \\
VoXtream2                                & \checkmark & 1.98 & 0.896 & 0.695 & 3.98 & \textbf{183} & 0.32 \\
CosyVoice3                               & \checkmark & 2.48 & \textbf{0.907} & 0.695 & 4.25 & 1553 & 0.34 \\
Faster-IndexTTS-2                        & \checkmark & 2.41 & 0.900 & 0.670 & 3.95 & 553 & \textbf{0.30} \\
Ours --- AR teacher (32-step)            & \ding{55} & 1.73 & 0.901 & 0.665 & 4.14 & - & - \\
Ours --- CD (4-step)                     & \checkmark & 2.05 & 0.887 & 0.631 & 3.82 & 286 & 0.31 \\
\textbf{Ours} --- DMD (4-step)           & \checkmark & 1.69 & 0.897 & 0.647 & 4.19 & 286 & 0.31 \\
\bottomrule
\end{tabular}}
\end{table}
\section{English objective results with both similarity metrics}
\label{app:ensim}

Both main-text short-form English tables (Tables~\ref{tab:tts_en} and~\ref{tab:vc_en}) headline WeSpeaker
(SimAM-ResNet100/VoxBlink2), our primary speaker-similarity metric. For completeness we repeat both
tables here with the two secondary variants, Resemblyzer (SIM$_{\text{Res}}$) and WavLM-large-SV
(SIM$_{\text{WavLM}}$); every other value is identical to the main text. The metrics rank speaker
similarity differently---most visibly on VC, where WavLM-SV places MeanVC-2 highest among streaming
systems while WeSpeaker and the listening study both place it lowest, and where the gap from our streaming
student to the X-VC leader is far smaller on WeSpeaker than on WavLM-SV. Both full tables also include
our intermediate-stage ablations---the $32$-step AR causal teacher and the CD student. The AR teacher is
chunk-causal but, at $32$ steps, slower than real time, so we mark it non-streaming and
report it only as an internal quality reference rather than a competing real-time system; the CD student
shares the DMD student's architecture, and therefore the same first-packet latency and RTF.

\begin{table}[t]
\centering\small
\setlength{\tabcolsep}{5pt}
\caption{\textbf{Short-form zero-shot streaming VC, Seed-TTS-eval English --- full table}. Companion to
main-text Table~\ref{tab:vc_en} with all three similarity metrics (WeSpeaker, Resemblyzer,
WavLM-large-SV), plus the internal AR-teacher and CD rows.}
\label{tab:vc_en_full}
\resizebox{\textwidth}{!}{%
\begin{tabular}{lcccccccc}
\toprule
System & Stream & WER\,$\downarrow$ & SIM$_{\text{WeSp}}$\,$\uparrow$ & SIM$_{\text{Res}}$\,$\uparrow$ & SIM$_{\text{WavLM}}$\,$\uparrow$ & UTMOS\,$\uparrow$ & FPL\,(ms)\,$\downarrow$ & RTF\,$\downarrow$ \\
\midrule
\multicolumn{9}{l}{\textit{Offline references}}\\
\midrule\
\textbf{Ours} --- offline              & \ding{55} & 3.11 & 0.653 & 0.848 & 0.476 & \textbf{4.11} & -- & -- \\
Vevo                                      & \ding{55} & 3.09 & \textbf{0.756} & \textbf{0.881} & 0.553 & 3.96 & -- & -- \\
\midrule
\multicolumn{9}{l}{\textit{Streaming systems}}\\
\midrule

Seed-VC                                  & \checkmark & 5.85 & 0.675 & 0.868 & 0.439 & 3.67 & 466 & 0.58 \\
X-VC                                     & \checkmark & 3.41 & \textbf{0.748} & \textbf{0.886} & 0.604 & 3.45 & 285 & 0.35 \\
MeanVC-2                                 & \checkmark & 6.20 & 0.632 & 0.824 & \textbf{0.616} & 3.79 & \textbf{211} & \textbf{0.20} \\
Ours --- AR teacher (32-step)            & \ding{55} & 3.20 & 0.660 & 0.830 & 0.464 & 3.88 & - & - \\
Ours --- CD (4-step)                     & \checkmark & 3.23 & 0.639 & 0.829 & 0.447 & 3.76 & 575 & 0.26 \\
\textbf{Ours} --- DMD (4-step)           & \checkmark & \textbf{3.09} & 0.670 & 0.842 & 0.465 & \textbf{3.90} & 575 & 0.26 \\
\bottomrule
\end{tabular}}
\end{table}
\begin{table}[h]
\centering\small
\setlength{\tabcolsep}{6pt}
\caption{Short-form zero-shot TTS, Seed-TTS-eval Mandarin. CER via
Paraformer-zh; SIM$_{\text{Res}}$ headline.}
\label{tab:tts_zh}
\begin{tabular}{lcccc}
\toprule
System & Stream & CER\,$\downarrow$ & SIM$_{\text{Res}}$\,$\uparrow$ & UTMOS\,$\uparrow$ \\
\midrule
\multicolumn{5}{l}{\textit{Offline references}}\\
\midrule
\textbf{Ours} --- offline & \ding{55}  & 1.76 & 0.895 & 3.42 \\
F5-TTS                                   & \ding{55}  & 1.54 & 0.906 & 3.34 \\
Faster-IndexTTS-2(offline)                        & \ding{55}  & 0.93 & 0.899 & 3.34 \\
IndexTTS-2.5                       & \ding{55}  & 2.02 & 0.901 & 3.32 \\
S1-mini                                   & \ding{55} & 1.01 & 0.850 & 3.82 \\
\midrule
\multicolumn{5}{l}{\textit{Streaming systems}}\\
\midrule
Qwen3-TTS                                & \checkmark & \textbf{0.91} & 0.881 & \textbf{3.94} \\
CosyVoice3                               & \checkmark & 1.41 & \textbf{0.910} & 3.65 \\
Faster-IndexTTS-2                        & \checkmark & 1.26 & 0.901 & 3.15 \\
Ours --- AR teacher (32-step)            & \ding{55} & 1.87 & 0.880 & 3.33 \\
Ours --- CD (4-step)                     & \checkmark & 2.77 & 0.872 & 2.97 \\
\textbf{Ours} --- DMD (4-step)           & \checkmark & 1.70 & 0.887 & 3.41 \\
\bottomrule
\end{tabular}
\end{table}
\begin{table}[h]
\centering\small
\setlength{\tabcolsep}{6pt}
\caption{Short-form zero-shot streaming VC, Seed-TTS-eval Mandarin.
Latency/RTF are the arch-shared values reported in Table~\ref{tab:latency}. Our CER is competitive with
X-VC; the WavLM-SV gap to baselines narrows relative to English.}
\label{tab:vc_zh}
\begin{tabular}{lcccc}
\toprule
System & Stream & CER\,$\downarrow$ & SIM$_{\text{Res}}$\,$\uparrow$ & UTMOS\,$\uparrow$ \\
\midrule
\multicolumn{5}{l}{\textit{Offline references}}\\
\midrule
\textbf{Ours} --- offline             & \ding{55}  & 2.82 & 0.854 & 3.36 \\
\midrule
\multicolumn{5}{l}{\textit{Streaming systems}}\\
\midrule
Seed-VC                                  & \checkmark & 4.28 & 0.838 & \textbf{3.43} \\
X-VC                                     & \checkmark & \textbf{2.73} & \textbf{0.894} & 2.49 \\
MeanVC-2                                 & \checkmark & 4.08 & 0.836 & 3.15 \\
Ours --- AR teacher (32-step)            & \ding{55} & 3.09 & 0.838 & 3.03 \\
Ours --- CD (4-step)                     & \checkmark & 3.29 & 0.833 & 2.91 \\
\textbf{Ours} --- DMD (4-step)           & \checkmark & 3.26 & 0.841 & 3.06 \\
\bottomrule
\end{tabular}
\end{table}
\section{Short-form Mandarin results}
\label{app:zh}

Tables~\ref{tab:tts_zh} and \ref{tab:vc_zh} report short-form Mandarin. On TTS the offline$\to$streaming
delta is now negligible in both languages---the streaming DMD student even edges the offline model on
CER/WER (Mandarin $1.70$ vs.\ $1.76$, English $1.69$ vs.\ $1.75$)---while VC incurs a modest $+0.44$ CER
cost. UTMOS is English-trained, so its Mandarin values are for cross-system ranking only. On TTS, our
streaming DMD has slightly higher CER but competitive similarity; the AR causal teacher is again the
stronger-but-slower point (causal but not real-time, so we mark it non-streaming) and consistency
distillation the weak link, matching the English ablation.

\section{Long-form results}
\label{app:longform}

The long-form benchmarks are designed to evaluate both output quality and speaker-similarity drift
during extended generation.

Table~\ref{tab:longform} gives English long-form \emph{TTS} identity consistency (windowed Resemblyzer
similarity: whole-clip SIM, early$\to$late halves, absolute drift, slope per minute, and long-form WER);
long-form \emph{VC} is carried by Figure~\ref{fig:rapr} in the main text, with the full
Mandarin table in Table~\ref{tab:vc_zh_long} below. TTS re-anchors per sentence and is flat by
construction.

\begin{table}[h]
\centering\small
\setlength{\tabcolsep}{6pt}
\caption{Long-form TTS identity consistency, English ($100$ passages $\sim$48\,s).}
\label{tab:longform}
\begin{tabular}{lccccc}
\toprule
System & whole-SIM$_{\text{Res}}$\,$\uparrow$ & early$\to$late & drift\,$\downarrow$ & slope/min\,$\uparrow$ & long-WER\,$\downarrow$ \\
\midrule
\textbf{CaDiS}          & 0.935 & 0.917$\to$0.909 & 0.009 & \textbf{$-$0.031} & 6.81 \\
CosyVoice3                     & \textbf{0.938} & 0.924$\to$0.916 & 0.008 & $-$0.034 & 7.24 \\
Qwen3-TTS                      & 0.885 & 0.875$\to$0.855 & 0.021 & $-$0.122 & \textbf{4.93} \\
\bottomrule
\end{tabular}
\end{table}
\begin{table}[h]
\centering\small
\setlength{\tabcolsep}{6pt}
\caption{Long-form TTS, Mandarin (sentence-concat $\sim$145\,s, $n{=}20$, enhanced targets). CER as \%.}
\label{tab:tts_zh_long}
\resizebox{\textwidth}{!}{%
\begin{tabular}{lcccccc}
\toprule
System & Stream & whole-SIM$_{\text{Res}}$\,$\uparrow$ & early$\to$late & drift\,$\downarrow$ & slope/min\,$\uparrow$ & long-CER\,$\downarrow$ \\
\midrule
\textbf{CaDiS}  & \checkmark & 0.922 & 0.899$\to$0.891 & \textbf{0.008} & \textbf{$-$0.004} & 8.4 \\
CosyVoice3             & \checkmark & \textbf{0.930} & 0.906$\to$0.894 & 0.013 & $-$0.006 & 7.4 \\
S1-mini                  & \ding{55} & 0.879 & 0.853$\to$0.844 & 0.010 & $-$0.006 & 4.8 \\
Qwen3-TTS              & \checkmark & 0.864 & 0.873$\to$0.794 & 0.079 & $-$0.056 & \textbf{4.6} \\
\bottomrule
\end{tabular}}
\end{table}

\begin{table}[h]
\centering\small
\setlength{\tabcolsep}{6pt}
\caption{Long-form VC, Mandarin (AISHELL-3 single-speaker src $\sim$122\,s, $n{=}20$, enhanced targets).
CER as \%.}
\label{tab:vc_zh_long}
\resizebox{\textwidth}{!}{%
\begin{tabular}{lcccccc}
\toprule
System & Stream & whole-SIM$_{\text{Res}}$\,$\uparrow$ & early$\to$late & drift\,$\downarrow$ & slope/min\,$\uparrow$ & long-CER\,$\downarrow$ \\
\midrule
\multicolumn{7}{l}{\textit{Offline references}}\\
\midrule
\textbf{Ours} --- offline          & \ding{55}  & 0.793 & 0.775$\to$0.769 & 0.006 & $-$0.004 & 7.2 \\
\midrule
\multicolumn{7}{l}{\textit{Streaming systems}}\\
\midrule
\textbf{CaDiS}                 & \checkmark & 0.869 & 0.837$\to$0.814 & 0.023 & $-$0.016 & 8.7 \\
X-VC                                  & \checkmark & \textbf{0.921} & 0.903$\to$0.891 & 0.012 & $-$0.009 & \textbf{7.1} \\
MeanVC-2                              & \checkmark & 0.872 & 0.851$\to$0.841 & 0.010 & $-$0.009 & 9.5 \\
Seed-VC                               & \checkmark & 0.752 & 0.726$\to$0.722 & \textbf{0.004} & $-$0.015 & 24.4 \\
\bottomrule
\end{tabular}}
\end{table}

\begin{table}[h]
\centering\small
\setlength{\tabcolsep}{5pt}
\caption{TTS on AniSpeech.}
\label{tab:anispeech_tts}
\resizebox{\textwidth}{!}{%
\begin{tabular}{lccccccc}
\toprule
System & Stream & WER\,$\downarrow$ & SIM$_{\text{WeSp}}$\,$\uparrow$ & \textit{SIM}$_{\text{WavLM}}$\,$\uparrow$ & UTMOS\,$\uparrow$ & FPL\,(ms)\,$\downarrow$ & RTF\,$\downarrow$ \\
\midrule
\multicolumn{8}{l}{\textit{Offline references}}\\
\midrule
F5-TTS                                   & \ding{55} & 3.34 & 0.748 & \textit{\textbf{0.628}} & 3.74 & -- & -- \\
IndexTTS-2.5                             & \ding{55} & 5.31 & \textbf{0.750} & \textit{0.588} & 3.67 & -- & -- \\
S1-mini                                   & \ding{55} & 3.07 & 0.604 & \textit{0.548} & \textbf{4.23} & -- & -- \\
\textbf{Ours} & \ding{55} & \textbf{2.64} & 0.740 & \textit{0.626} & 3.89 & -- & -- \\
\midrule
\multicolumn{8}{l}{\textit{Streaming systems}}\\
\midrule
Qwen3-TTS                          & \checkmark & \textbf{2.65} & 0.668 & \textit{0.601} & 3.98 & 334 & 0.39 \\
VoXtream2                                & \checkmark & 5.24 & 0.694 & \textit{0.631} & 3.37 & \textbf{183} & 0.32 \\
CosyVoice3                         & \checkmark & 3.06 & \textbf{0.771} & \textit{\textbf{0.645}} & \textbf{4.04} & 923 & 0.34 \\
Faster-IndexTTS-2                        & \checkmark & 5.36 & 0.761 & \textit{0.638} & 3.51 & 553 & \textbf{0.30} \\
\textbf{CaDiS}           & \checkmark & 3.12 & 0.690 & \textit{0.590} & 3.67 & 286 & 0.31 \\
\bottomrule
\end{tabular}}
\end{table}

\begin{table}[h]
\centering\small
\setlength{\tabcolsep}{5pt}
\caption{Streaming cross-speaker VC on AniSpeech (additional anime benchmark, OOD; enhanced targets).
The two \emph{Ours} rows isolate the direct-metric addition (Appendix~\ref{app:vcablate}). $^{a}$MeanVC-2 is Mandarin-only (English is OOD for it).}
\label{tab:anispeech_vc}
\resizebox{\textwidth}{!}{%
\begin{tabular}{lccccccc}
\toprule
System & Stream & WER\,$\downarrow$ & SIM$_{\text{WeSp}}$\,$\uparrow$ & \textit{SIM}$_{\text{WavLM}}$\,$\uparrow$ & UTMOS\,$\uparrow$ & FPL\,(ms)\,$\downarrow$ & RTF\,$\downarrow$ \\
\midrule
\multicolumn{8}{l}{\textit{Offline references}}\\
\midrule
\textbf{Ours} --- offline              & \ding{55} & 5.58 & 0.596 & \textit{0.408} & \textbf{3.47} & -- & -- \\
\midrule
\multicolumn{8}{l}{\textit{Streaming systems}}\\
\midrule
Seed-VC                                   & \checkmark & 15.82 & 0.663 & \textit{0.440} & 3.06 & 466 & 0.58 \\
X-VC                             & \checkmark & \textbf{4.92} & \textbf{0.674} & \textit{0.525} & 2.88 & 285 & 0.35 \\
MeanVC-2$^{a}$                            & \checkmark & 9.04 & 0.596 & \textit{\textbf{0.588}} & 3.46 & \textbf{211} & \textbf{0.20} \\
\textbf{CaDiS}      & \checkmark & 5.30 & 0.576 & \textit{0.399} & 3.37 & 575 & 0.26 \\
\bottomrule
\end{tabular}}
\end{table}

\begin{table}[h]
\centering\small
\setlength{\tabcolsep}{4pt}
\caption{\textbf{TTS on Anim400K}. Objective metrics
on the full benchmark plus subjective MOS from a gender-balanced listening study (Prolific; protocol in
Appendix~\ref{app:subjective}). All systems in
the table were evaluated together in the same collection. Streaming$-$offline (Ours): $\Delta_{\text{NMOS}}{=}{-}0.11$,
$\Delta_{\text{QMOS}}{=}{-}0.23$, $\Delta_{\text{SMOS}}{=}{-}0.21$. The $95\%$ CIs for QMOS and SMOS
exclude $0$, whereas the NMOS CI includes $0$ ($[-0.22,+0.00]$); on this
out-of-domain benchmark the streaming student does not fully close the gap to the offline teacher.}
\label{tab:anim400k_tts}
\resizebox{\textwidth}{!}{%
\begin{tabular}{lcccc|ccc}
\toprule
& & \multicolumn{3}{c|}{Objective} & \multicolumn{3}{c}{Subjective MOS} \\
\cmidrule(lr){3-5}\cmidrule(lr){6-8}
System & Stream & WER\,$\downarrow$ & SIM$_{\text{WeSp}}$\,$\uparrow$ & UTMOS\,$\uparrow$ & NMOS\,$\uparrow$ & QMOS\,$\uparrow$ & SMOS\,$\uparrow$ \\
\midrule
\multicolumn{8}{l}{\textit{Offline references}}\\
\midrule
\textbf{Ours} --- offline & \ding{55} & \textbf{1.14} & \textbf{0.788} & 4.32 & \textbf{3.66} & \textbf{4.10} & \textbf{3.54} \\
F5-TTS                                   & \ding{55} & 1.79 & 0.781 & 4.21 & 3.41 & 3.91 & 3.28 \\
IndexTTS-2.5                             & \ding{55} & 3.21 & 0.765 & 4.05 & 3.51 & 3.89 & 3.39 \\
S1-mini                                   & \ding{55} & 1.16 & 0.675 & \textbf{4.41} & 3.36 & 4.02 & 3.10 \\
\midrule
\multicolumn{8}{l}{\textit{Streaming systems}}\\
\midrule
Qwen3-TTS                                & \checkmark & \textbf{0.90} & 0.712 & 4.23 & \textbf{3.73} & \textbf{4.08} & 3.24 \\
VoXtream2                                & \checkmark & 1.51 & 0.762 & 4.12 & 3.39 & 3.91 & 3.35 \\
CosyVoice3                               & \checkmark & 1.39 & \textbf{0.784} & \textbf{4.33} & 3.60 & \textbf{4.08} & \textbf{3.47} \\
Faster-IndexTTS-2                        & \checkmark & 3.29 & 0.773 & 3.99 & 3.42 & 3.73 & 3.26 \\
\textbf{CaDiS}     & \checkmark & 1.17 & 0.753 & 4.20 & 3.55 & 3.87 & 3.33 \\
\bottomrule
\end{tabular}}
\end{table}
Tables~\ref{tab:tts_zh_long} and \ref{tab:vc_zh_long} give Mandarin long-form drift. As in English,
our TTS keeps the flattest slope at a bounded window, and VC drifts gracefully with content held by the
live source.

\section{Out-of-domain anime results}
\label{app:ood}

Anim400K is our \textbf{main anime benchmark}: its cross-speaker VC results are promoted to the main
text (Table~\ref{tab:anim400k_vc}) and its TTS results appear in Table~\ref{tab:anim400k_tts}. We
additionally report \textbf{AniSpeech} (Tables~\ref{tab:anispeech_tts}--\ref{tab:anispeech_vc}) as a
second anime benchmark for breadth. On both, SIM$_{\text{WeSp}}$ (SimAM-ResNet100/VoxBlink2) is the
headline speaker metric, and both anime source and target are enhanced. Before constructing the evaluation pairs, we apply model-independent audio-quality, single-speaker-purity, and transcript-consistency filtering. The transcript-consistency
check removes segments containing approximately $1$\,s of neighboring speech caused by imperfect utterance boundaries. Filtering is completed before system evaluation, and no scoring-time trimming is applied.

\section{TTS positional encoding and sentence-length conditioning}
\label{app:tts_positional_ablation}

We compare two ways of positional encoding in full-text-ready streaming TTS:
retaining the offline teacher's normalized positional encoding, and replacing it with RoPE plus
explicit length conditioning. For the latter, we train an additional streaming TTS model through the
same three-stage pipeline and provide the total number of acoustic latent frames,
\texttt{num\_frames}, through adaptive layer normalization. During training, this length is derived
from the training audio; during inference, it is supplied by the duration estimator.

\begin{table}[tbp]
\centering\small
\setlength{\tabcolsep}{5pt}
\caption{\textbf{TTS positional encoding and length-conditioning comparison},
Seed-TTS-eval English.}
\label{tab:tts_positional_ablation}
\resizebox{\textwidth}{!}{%
\begin{tabular}{lccccc}
\toprule
Method & WER\,(\%)\,$\downarrow$ & SIM$_{\text{WeSp}}$\,$\uparrow$
& UTMOS\,$\uparrow$ & FPL\,(ms)\,$\downarrow$ & RTF\,$\downarrow$ \\
\midrule
Normalized positional encoding
& \textbf{1.69} & \textbf{0.785} & 4.19 & \textbf{286} & \textbf{0.31} \\
RoPE + \texttt{num\_frames}
& 2.34 & 0.783 & \textbf{4.27} & 306 & 0.34 \\
\bottomrule
\end{tabular}}
\end{table}

Table~\ref{tab:tts_positional_ablation} shows a trade-off rather than uniform superiority.
The retained normalized positional encoding yields lower WER and lower reported FPL/RTF, while the
RoPE variant yields higher UTMOS; the speaker-similarity results are close. These results
suggest that explicit sentence-length conditioning is a viable alternative to incorporating length
information through normalized positional encoding. 

\paragraph{Duration estimation and input condition.}
The duration estimator is a reference-rate heuristic, not a separately trained neural network.
We derive the average number of acoustic latent frames per IPA token from the reference audio and
its transcript, then use this rate to estimate the target sentence duration, with adjustments for
punctuation and non-speech events. The complete target sentence is available before synthesis begins.
The reported long-form TTS system resets at sentence boundaries.
\section{VC direct-metric ablation}
\label{app:vcablate}

The direct-metric term is flat on raw speaker cosine (WeSpeaker $0.576\to0.571$, within noise) but
improves intelligibility and naturalness (WER $5.30\to5.15$, UTMOS $3.37\to3.41$). In other words, the
direct-metric term buys WER/UTMOS, not speaker cosine.

\section{VC content-representation ablation: native \texorpdfstring{$40\hz$}{40 Hz} vs.\ \texorpdfstring{$50\hz$}{50 Hz}}
\label{app:contentvec}

Matching the content rate to the acoustic-latent rate is a design choice, not just an implementation
convenience. To separate the effect of the \emph{rate} from the effect of making the encoder
\emph{causal and streamable}, we probe three ContentVec encoders \emph{intrinsically}---without
retraining the VC system---on LibriSpeech test-clean ($40$ speakers $\times\,12$ utterances): the
pretrained $50\hz$ teacher (bidirectional reference); a \emph{strided} $40\hz$ encoder that only
relocates the convolutional stride to emit $40\hz$ (weights copied from the teacher, no training, still
bidirectional); and the deployed \emph{native $40\hz$} encoder, distilled from the teacher with causal
convolutions and block-causal self-attention. We measure content (phone error rate from a shared rate-agnostic CTC probe, so $40$ vs.\
$50\hz$ are comparable), speaker leakage (layer-12 linear speaker-classification accuracy; lower is more
invariant), pitch leakage (ridge F0 $R^2$; lower is more invariant), and cosine to the teacher
representation (Table~\ref{tab:contentvec}).

The decomposition is clean. Changing the rate alone (teacher50$\to$strided40) is nearly free on content
($+0.22$\,pt PER) and leaves speaker and pitch leakage flat or slightly lower; the larger cost
($+1.86$\,pt PER, teacher-cos $0.911\!\to\!0.847$) comes from making the encoder causal and streamable,
not from the rate change. The native $40\hz$ representation thus preserves content and disentanglement
while removing the condition$\leftrightarrow$latent rate mismatch and enabling direct additive
conditioning---confirming that the $40\hz$ design is functionally motivated rather than a convenience.

\begin{table}[h]
\centering\small
\setlength{\tabcolsep}{6pt}
\caption{\textbf{Intrinsic content-representation probes} for the ContentVec encoder (source-rate,
no VC retraining; LibriSpeech test-clean, $40$ spk $\times\,12$ utt). PER from a shared rate-agnostic
CTC probe (content); spk-acc is layer-12 linear speaker classification (leakage, chance $0.025$);
F0\,$R^2$ is ridge pitch prediction (leakage); teacher-cos is cosine to the $50\hz$ teacher.
\emph{teacher50}$\to$\emph{strided40} isolates the pure rate change; \emph{strided40}$\to$\emph{native40}
isolates the causal/streaming-distillation cost.}
\label{tab:contentvec}
\begin{tabular}{lcccc}
\toprule
Encoder & PER\,$\downarrow$ & spk-acc\,$\downarrow$ & F0\,$R^2$\,$\downarrow$ & teacher-cos\,$\uparrow$ \\
\midrule
teacher50 ($50\hz$ bidirectional, reference) & \textbf{5.39} & 0.690 & 0.356 & 1.000 \\
strided40 ($40\hz$, no training)             & 5.61 & \textbf{0.661} & \textbf{0.346} & 0.911 \\
native40 (distilled causal, deployed)        & 7.47 & 0.674 & 0.358 & 0.847 \\
\bottomrule
\end{tabular}
\end{table}

\section{RAPR window ablation: short-form quality, long-form drift, and attention}
\label{app:rapr}
\subsection{Fixed-history positional re-anchoring control}
\label{app:rapr_fixed_history}

Varying the rolling-history window changes both positional behavior and the acoustic context
available to the model, so a window-length sweep alone does not isolate positional re-encoding.
We therefore compare bounded-history decoding with monotonically increasing positions against
RAPR's positional re-anchoring while fixing the speaker reference, retained-history length,
chunk sizes, and sampling budget.

Both settings use a $1$\,s retained-history window, $400$\,ms generator and ContentVec chunks,
and four sampling steps. The monotonic-position control also undergoes Stage-3 distribution
matching refinement. Thus, the comparison tests positional re-anchoring at a fixed history budget.

\begin{table}[tbp]
\centering\small
\setlength{\tabcolsep}{6pt}
\caption{\textbf{Fixed-history positional control, short-form English VC}.}
\label{tab:rapr_fixed_short}
\begin{tabular}{lccc}
\toprule
Positional policy & WER\,(\%)\,$\downarrow$
& SIM$_{\text{WeSp}}$\,$\uparrow$ & UTMOS\,$\uparrow$ \\
\midrule
Monotonic, $1$\,s history & 2.90 & 0.674 & \textbf{3.94} \\
RAPR, $1$\,s history     & \textbf{2.83} & \textbf{0.675} & 3.93 \\
\bottomrule
\end{tabular}
\end{table}

On short-form VC, the evaluation results are close
(Table~\ref{tab:rapr_fixed_short}). The long-form comparison separates the two positional policies more clearly
(Table~\ref{tab:rapr_fixed_long}). With monotonic positions, mean target-speaker similarity falls
from $0.619$ to $0.125$, despite Stage-3 refinement and the same bounded history.
With RAPR, it remains nearly unchanged ($0.699\to0.696$), and full-clip WER is lower
($9.18\%$ versus $15.39\%$). These results support positional re-anchoring for long-form VC
under the evaluated fixed-history setting, rather than a general short-form quality advantage.

\begin{table}[tbp]
\centering\small
\setlength{\tabcolsep}{5pt}
\caption{\textbf{Fixed-history positional control, long-form English VC}.
Early and late target SIM are WeSpeaker similarities averaged over non-overlapping $10$\,s
windows whose centers lie within the first and last $30$\,s of each clip, respectively.}
\label{tab:rapr_fixed_long}
\resizebox{\textwidth}{!}{%
\begin{tabular}{lcccc}
\toprule
Positional policy & Full-clip WER\,(\%)\,$\downarrow$
& Early target SIM\,$\uparrow$ & Late target SIM\,$\uparrow$
& Late $-$ early SIM \\
\midrule
Monotonic, $1$\,s history & 15.39 & 0.619 & 0.125 & $-0.494$ \\
RAPR, $1$\,s history     & \textbf{9.18} & \textbf{0.699}
& \textbf{0.696} & $-0.003$ \\
\bottomrule
\end{tabular}}
\end{table}

\subsection{Rolling-window and attention analyses}
We ablate reference-anchored positional re-encoding by training four DMD students that differ
\emph{only} in the rolling clean-history window (train${=}$test, $0.5/1/4/10$\,s; the $\sim$10\,s speaker
reference and everything else are held fixed), plus an unbounded (RAPR-off) arm evaluated on the $1$\,s
model. These two ablation tables, with the long-form sweep in the main text
(Figure~\ref{fig:rapr}(a)) shows that a short window is enough for streaming generation in our setting.

\emph{Short-form quality} (Table~\ref{tab:rapr_short}). On both English and Mandarin Seed-TTS-eval VC,
speaker similarity (Resemblyzer and WeSpeaker) falls \emph{monotonically} as the window grows, while
intelligibility is flat (EN) or improves only $\sim$0.4\,pt (ZH); the $1$\,s row reproduces the deployed
anchor. The only thing a longer window buys is a small ZH content gain, at a similarity and cost penalty.

\emph{Long-form drift} (main text, Figure~\ref{fig:rapr}(a); full Mandarin table in
Table~\ref{tab:vc_zh_long}). Over minute-long rollouts, bounded windows all hold and \emph{smaller bounds
drift less, monotonically}, while disabling RAPR (unbounded) collapses identity. Bounding is essential
and a short bound is best---the opposite of the ``longer context helps'' intuition.

\emph{Mechanism} (Figure~\ref{fig:rapr}(b), Table~\ref{tab:rapr_attn}). Recovering the decoder
self-attention over the same $20$ rollouts shows the pinned prompt-sink's share of attention falls
monotonically with window size ($0.467\!\to\!0.121\!\to\!0.011$) and tracks the drift ranking
one-for-one: the RAPR window sets how large a fraction of the KV cache the fixed speaker anchor occupies.
A short window lets it dominate; an unbounded window lets accumulated (and drifting) source-history wash
it out. This is the mechanism behind the VC identity claim, measured rather than argued.

\begin{table}[h]
\centering\small
\setlength{\tabcolsep}{5pt}
\caption{\textbf{RAPR rolling-window sweep, short-form Seed-TTS-eval VC} (4 steps, enhanced sources;
speaker reference fixed at $\sim$10\,s; only the train-matched rolling window varies). Speaker
similarity (Res${=}$Resemblyzer headline, WeSp${=}$WeSpeaker) falls monotonically as the window grows on
both languages; the $1$\,s row ($\star$, deployed) reproduces the anchor exactly.}
\label{tab:rapr_short}
\begin{tabular}{lcccc|cccc}
\toprule
& \multicolumn{4}{c|}{English} & \multicolumn{4}{c}{Mandarin} \\
Window & WER\,$\downarrow$ & SIM$_{\text{Res}}$\,$\uparrow$ & SIM$_{\text{WeSp}}$\,$\uparrow$ & UTMOS & CER\,$\downarrow$ & SIM$_{\text{Res}}$\,$\uparrow$ & SIM$_{\text{WeSp}}$\,$\uparrow$ & UTMOS \\
\midrule
$0.5$\,s        & 3.33 & \textbf{0.849} & \textbf{0.681} & 3.90 & 3.45 & \textbf{0.847} & \textbf{0.776} & 3.07 \\
$1$\,s$^{\star}$ & \textbf{3.09} & 0.842 & 0.670 & 3.90 & 3.26 & 0.841 & 0.772 & 3.06 \\
$4$\,s          & 3.16 & 0.834 & 0.657 & 3.88 & 3.05 & 0.828 & 0.758 & 3.06 \\
$10$\,s         & 3.16 & 0.835 & 0.657 & 3.88 & \textbf{3.01} & 0.828 & 0.757 & 3.07 \\
\bottomrule
\end{tabular}
\end{table}

\begin{table}[h]
\centering\small
\setlength{\tabcolsep}{6pt}
\caption{\textbf{RAPR attention mechanism} (decoder self-attention recovered from 20 cross-speaker
rollouts ${\sim}121$\,s, averaged over steps/heads/queries/$20$ layers and clips; late half
$t\ge t_{\text{last}}/2$). The three key regions partition the attention mass (rows sum to $1$). Mass on
the pinned prompt-sink falls monotonically as the window grows---displaced by the accumulating
clean-history region---and tracks the long-form drift ranking (Figure~\ref{fig:rapr}(a))
one-for-one.}
\label{tab:rapr_attn}
\begin{tabular}{lccc}
\toprule
Window & prompt-sink\,$\uparrow$ & clean-history & current \\
\midrule
$0.5$\,s & \textbf{0.467} & 0.160 & 0.373 \\
$1$\,s$^{\star}$ & 0.417 & 0.226 & 0.356 \\
$10$\,s & 0.121 & 0.594 & 0.286 \\
Unbounded & 0.011 & 0.755 & 0.234 \\
\bottomrule
\end{tabular}
\end{table}

\section{Sampling-step budget}
\label{app:steps}

We sweep student NFEs ($1/2/4$ vs.\ the $32$-step teacher). Four
steps is the knee of the quality--latency curve: TTS intelligibility and naturalness reach the offline
ceiling while RTF stays at $0.31$; below four steps UTMOS degrades faster than WER, and speaker
similarity is the last to fall. For streaming VC, Table~\ref{tab:vc_steps} sweeps $1/2/4/8$ steps on the
deployed VC DMD student (EN Seed-TTS-eval; everything else fixed). The same
knee appears: WER and speaker similarity (Resemblyzer) both peak at four steps and UTMOS reaches within
$0.05$ of its ceiling, while going to eight steps \emph{regresses} WER ($3.09\to3.24$) and Resemblyzer
($0.842\to0.838$)---over-denoising---for a doubled per-chunk cost. One step is fastest but sacrifices
naturalness (UTMOS $3.33$).

\begin{table}[h]
\centering\small
\setlength{\tabcolsep}{6pt}
\caption{Streaming VC sampling-step budget on the deployed VC DMD student (EN Seed-TTS-eval, enhanced
sources; $\sim$1\,s rolling context, $\sim$10\,s reference). Only the per-chunk step count varies; the
four-step row reproduces the deployed system exactly. SIM$_{\text{Res}}$ (Resemblyzer) is the headline
and SIM$_{\text{WeSp}}$ (WeSpeaker) secondary. Four steps ($\star$) is the deployed default; deployed
latency is reported in Table~\ref{tab:latency}.}
\label{tab:vc_steps}
\begin{tabular}{ccccc}
\toprule
Steps & WER\,$\downarrow$ & SIM$_{\text{Res}}$\,$\uparrow$ & SIM$_{\text{WeSp}}$\,$\uparrow$ & UTMOS\,$\uparrow$ \\
\midrule
1        & 3.57 & 0.831 & 0.629 & 3.33 \\
2        & 3.23 & 0.841 & 0.654 & 3.76 \\
4\,$\star$ & \textbf{3.09} & \textbf{0.842} & 0.670 & 3.90 \\
8        & 3.24 & 0.838 & \textbf{0.672} & \textbf{3.95} \\
\bottomrule
\end{tabular}
\end{table}
\section{VC chunk-size quality--latency trade-off}
\label{app:vc_chunk_size}

We evaluate inference-time chunk-size changes using the same streaming VC checkpoint, trained
with $400$\,ms chunks. We vary the generator and ContentVec chunk sizes together across
$100$, $200$, and $400$\,ms, while keeping four sampling steps and a $1$\,s RAPR history.
No retraining is performed for the smaller chunk sizes.

\begin{table}[tbp]
\centering\small
\setlength{\tabcolsep}{5pt}
\caption{\textbf{Inference-time VC chunk-size sweep}, Seed-TTS-eval English.
Generator and ContentVec chunk sizes are varied together using the same checkpoint.}
\label{tab:vc_chunk_size}
\begin{tabular}{cccccc}
\toprule
Chunk size & WER\,(\%)\,$\downarrow$ & SIM$_{\text{WeSp}}$\,$\uparrow$
& UTMOS\,$\uparrow$ & FPL\,(ms)\,$\downarrow$
& RTF\,$\downarrow$ \\
\midrule
$100$\,ms & 3.42 & 0.671 & 3.72 & \textbf{272} & 1.10 \\
$200$\,ms & 3.27 & \textbf{0.675} & 3.85 & 373 & 0.59 \\
$400$\,ms & \textbf{2.85} & \textbf{0.675} & \textbf{3.93}
& 576 & \textbf{0.33} \\
\bottomrule
\end{tabular}
\end{table}

Table~\ref{tab:vc_chunk_size} shows a quality--latency trade-off. Smaller chunks reduce startup
latency, while WER and UTMOS worsen and speaker similarity changes relatively little.
At $100$\,ms, FPL approaches the reported X-VC operating point ($272$ versus $285$\,ms), but
processing RTF rises to $1.10$. This setting therefore does not establish sustained real-time
conversion at matched startup latency. The $200$\,ms setting provides an intermediate operating
point, reducing FPL to $373$\,ms with processing RTF $0.59$, at some intelligibility and naturalness
cost relative to the $400$\,ms variant.
\section{Subjective evaluation protocol}
\label{app:subjective}

Because neural MOS predictors saturate on strong systems, we validate naturalness with a human
listening study. Rather than one omnibus questionnaire, we decompose each evaluation into
\emph{separate rating tasks}---\textbf{two} for TTS and \textbf{three} for VC---each a self-contained
Prolific instrument with its own instructions, its own on-screen reference condition, and its own
randomized trial pages. Within a task, all systems for that task--language--benchmark are rated
together and a given rater/item pair never crosses tasks. The two shared tasks are (i)~\emph{naturalness
and quality}, which scores the synthesized recording \emph{in isolation} on five-point naturalness
(NMOS) and quality/artifact (QMOS) scales, and (ii)~\emph{target-speaker similarity} (SMOS), which
plays the recording beside a reference timbre---the prompt for TTS, the target for VC. VC adds a third
task, (iii)~\emph{source preservation}, which plays the converted output beside the \emph{source}
recording and scores rhythm/timing (\textsc{Rhy}) and expression/emotion (\textsc{Exp}); lexical content
is already covered by objective WER. Every rating is a five-point absolute score. Raters are Prolific
English speakers (English evaluations) or native/fluent-Mandarin speakers (Mandarin evaluations), item-blocked
with two votes per cell; attention-failed assignments are dropped whole ($6$--$18\%$ per English family,
$17$--$28\%$ per Mandarin family), means carry participant-cluster bootstrap $95\%$ CIs. Listening sets are gender-balanced, specifically 100 male utterances and 100 female utterances for TTS experiments, and 50 male-to-female conversion, 50 female-to-male conversion, 50 male-to-male conversion, and 50 female-to-female conversion for VC experiments. The results are reported as follows:
the English Seed-TTS-eval results are in the main text (TTS Table~\ref{tab:tts_en},
Section~\ref{sec:tts_results}; VC Table~\ref{tab:vc_en}, Section~\ref{sec:vc_results}); the
out-of-domain Anim400K results are folded into Appendix~\ref{app:ood} (TTS Table~\ref{tab:anim400k_tts},
VC Table~\ref{tab:vc_subj_anim}); and the Mandarin (Seed-TTS-eval) results are reported below
(Tables~\ref{tab:tts_subj_zh}--\ref{tab:vc_subj_zh}).

\begin{figure}[p]
\centering
\includegraphics[width=\linewidth]{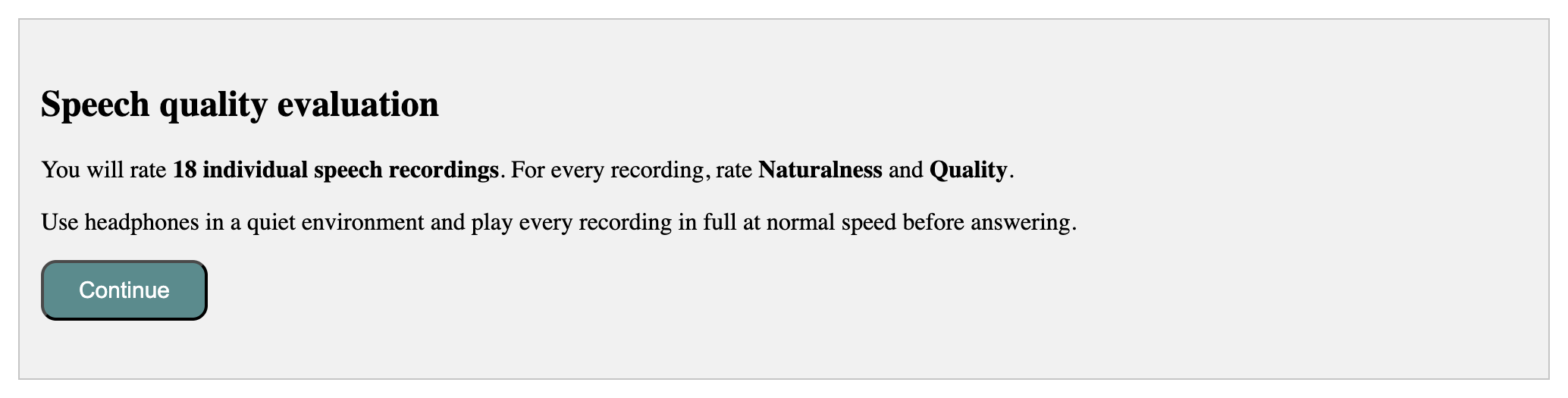}\\[4pt]
\includegraphics[width=\linewidth]{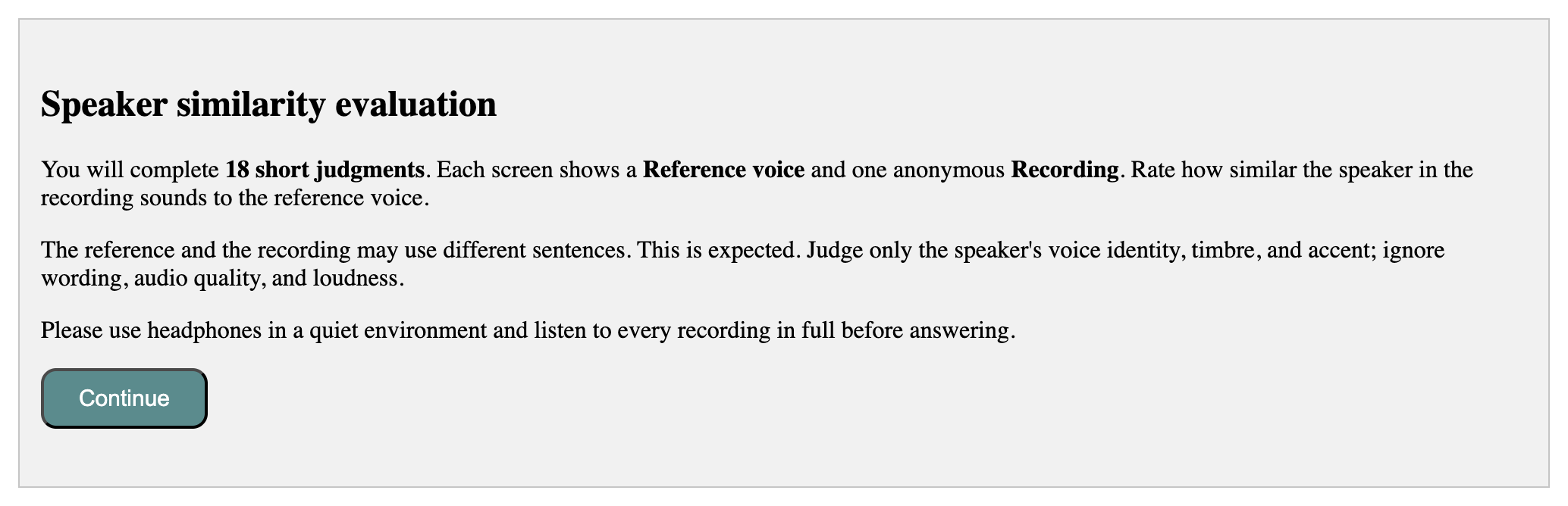}\\[4pt]
\includegraphics[width=\linewidth]{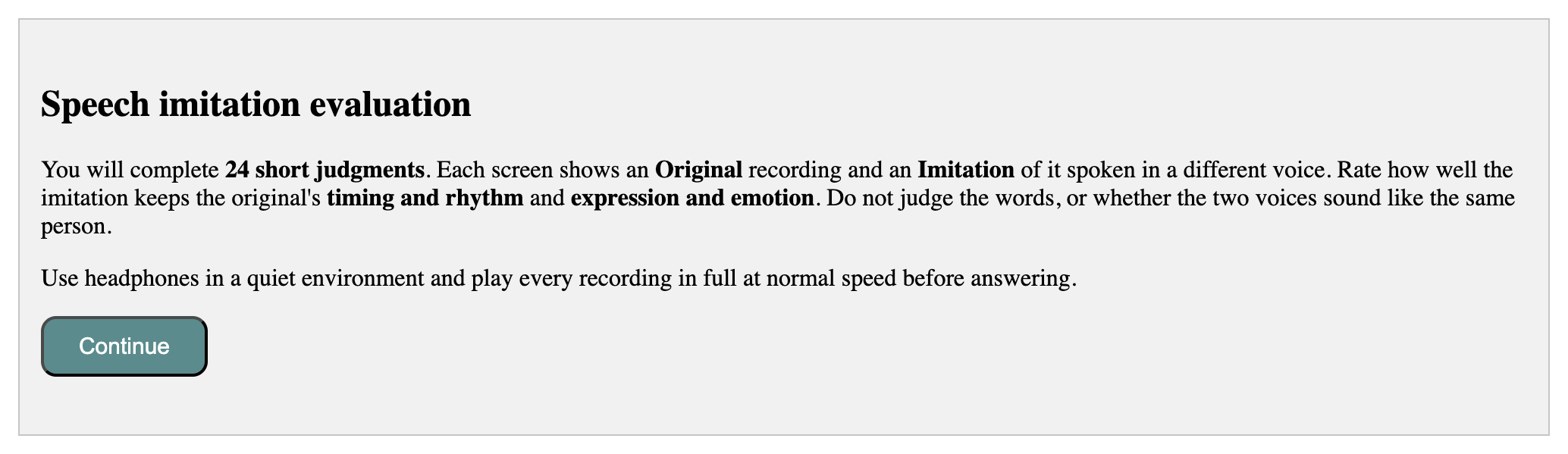}
\caption{\textbf{Participant instructions for the English listening tasks} (welcome screens). \emph{Top:}
the naturalness/quality task rates each recording on its own. \emph{Middle:} the similarity task shows a
reference voice beside the recording and asks the rater to attend only to voice identity, timbre, and
accent. \emph{Bottom:} the VC-only source-preservation (``imitation'') task asks only about timing/rhythm
and expression/emotion, not the words or the speaker. TTS uses the top two instruments ($18$ recordings
per assignment); VC uses all three ($24$ per assignment).}
\label{fig:subj_instr}
\end{figure}

\begin{figure}[p]
\centering
\begin{subfigure}{\linewidth}\centering
  \includegraphics[width=\linewidth]{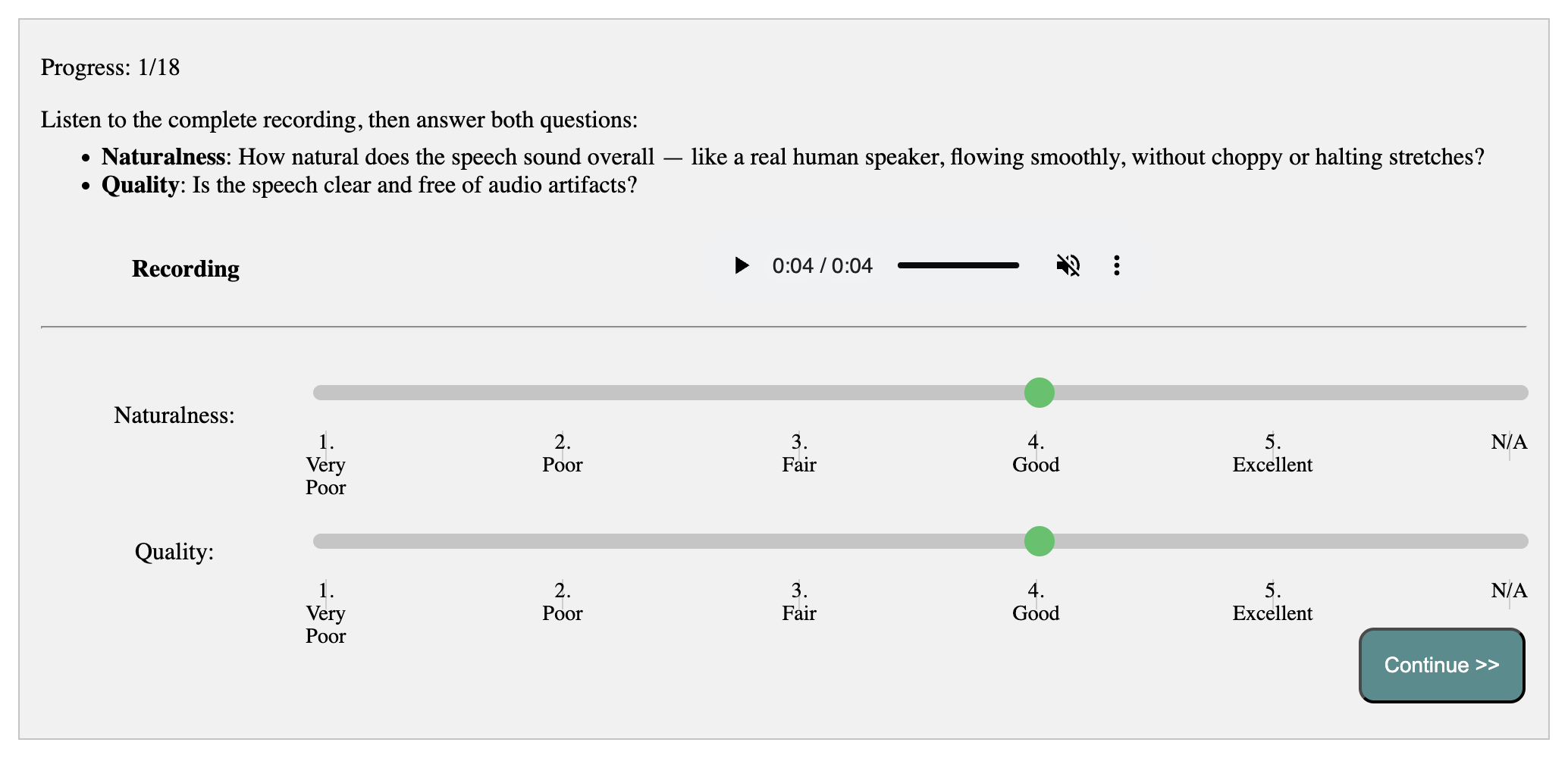}
  \caption{Naturalness \& quality (NMOS/QMOS): the recording is rated on its own, with no reference voice.}
  \label{fig:subj_tts_q}
\end{subfigure}\\[8pt]
\begin{subfigure}{\linewidth}\centering
  \includegraphics[width=\linewidth]{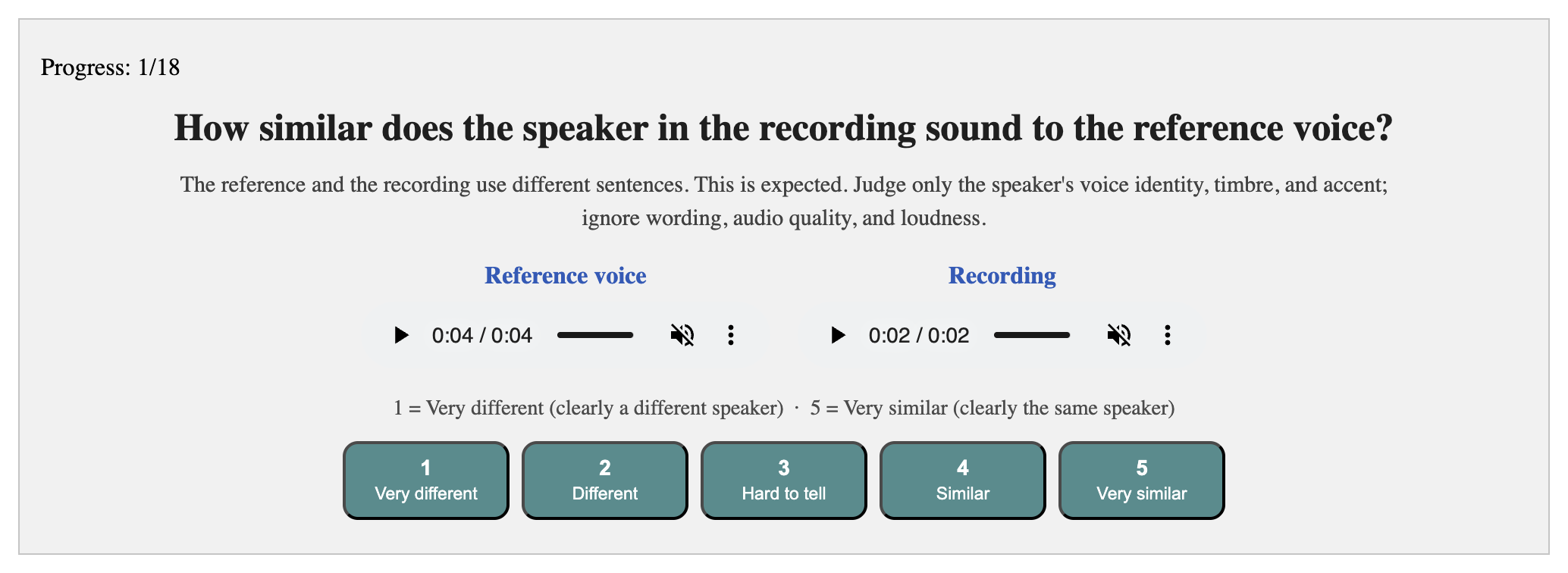}
  \caption{Target-speaker similarity (SMOS): the prompt is shown as the ``Reference voice'' beside the recording.}
  \label{fig:subj_tts_s}
\end{subfigure}
\caption{\textbf{English TTS listening-study interface: two tasks.} The prompt timbre appears only in the
similarity task (b), never in the quality task (a), so naturalness/quality scores cannot be contaminated
by the reference clip.}
\label{fig:subj_tts}
\end{figure}

\begin{figure}[p]
\centering
\begin{subfigure}{\linewidth}\centering
  \includegraphics[width=0.72\linewidth]{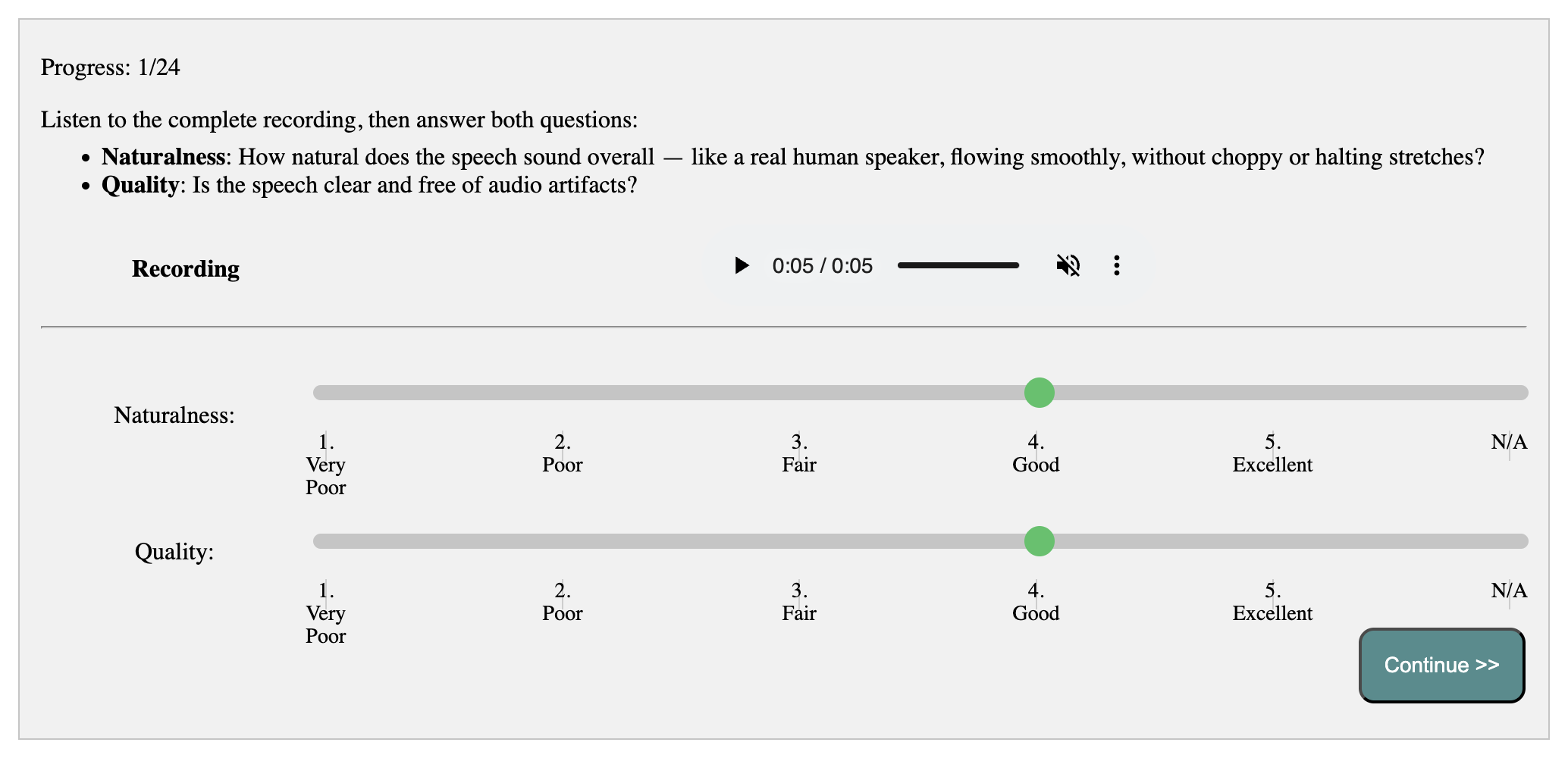}
  \caption{Naturalness \& quality (NMOS/QMOS): the converted output is rated on its own, no reference.}
  \label{fig:subj_vc_q}
\end{subfigure}\\[8pt]
\begin{subfigure}{\linewidth}\centering
  \includegraphics[width=0.72\linewidth]{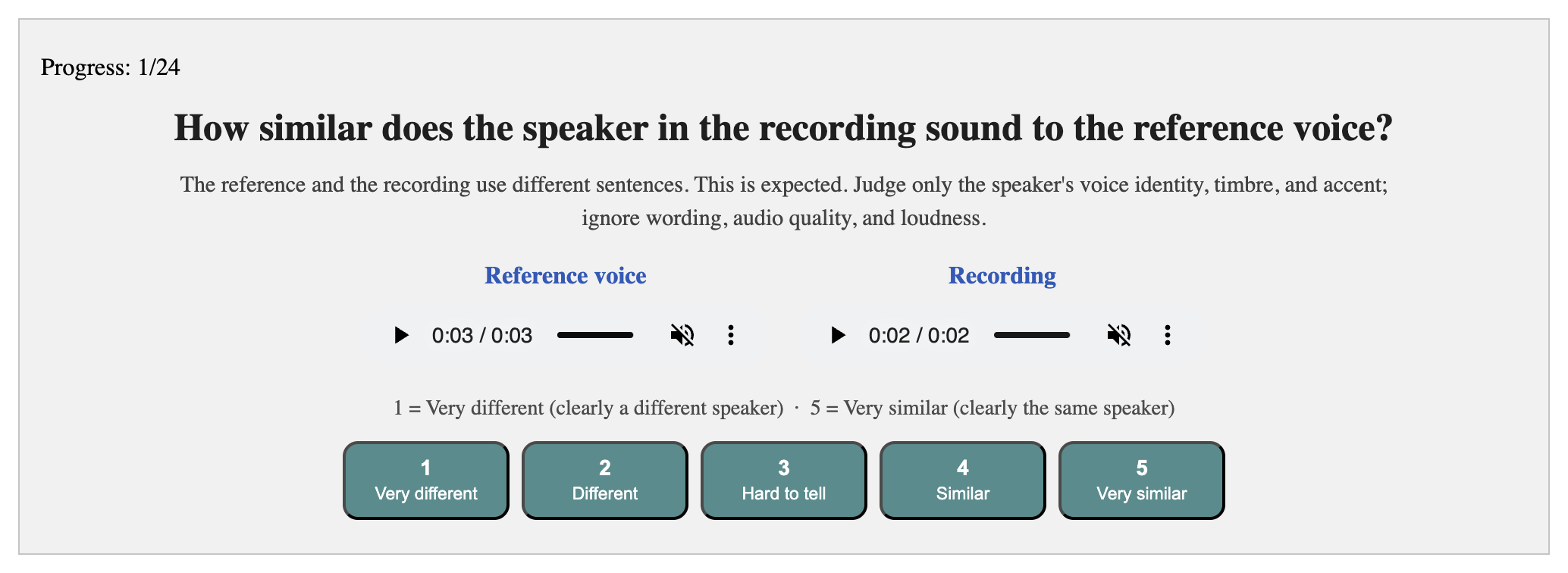}
  \caption{Target-speaker similarity (SMOS): the conversion target is the ``Reference voice.''}
  \label{fig:subj_vc_s}
\end{subfigure}\\[8pt]
\begin{subfigure}{\linewidth}\centering
  \includegraphics[width=0.72\linewidth]{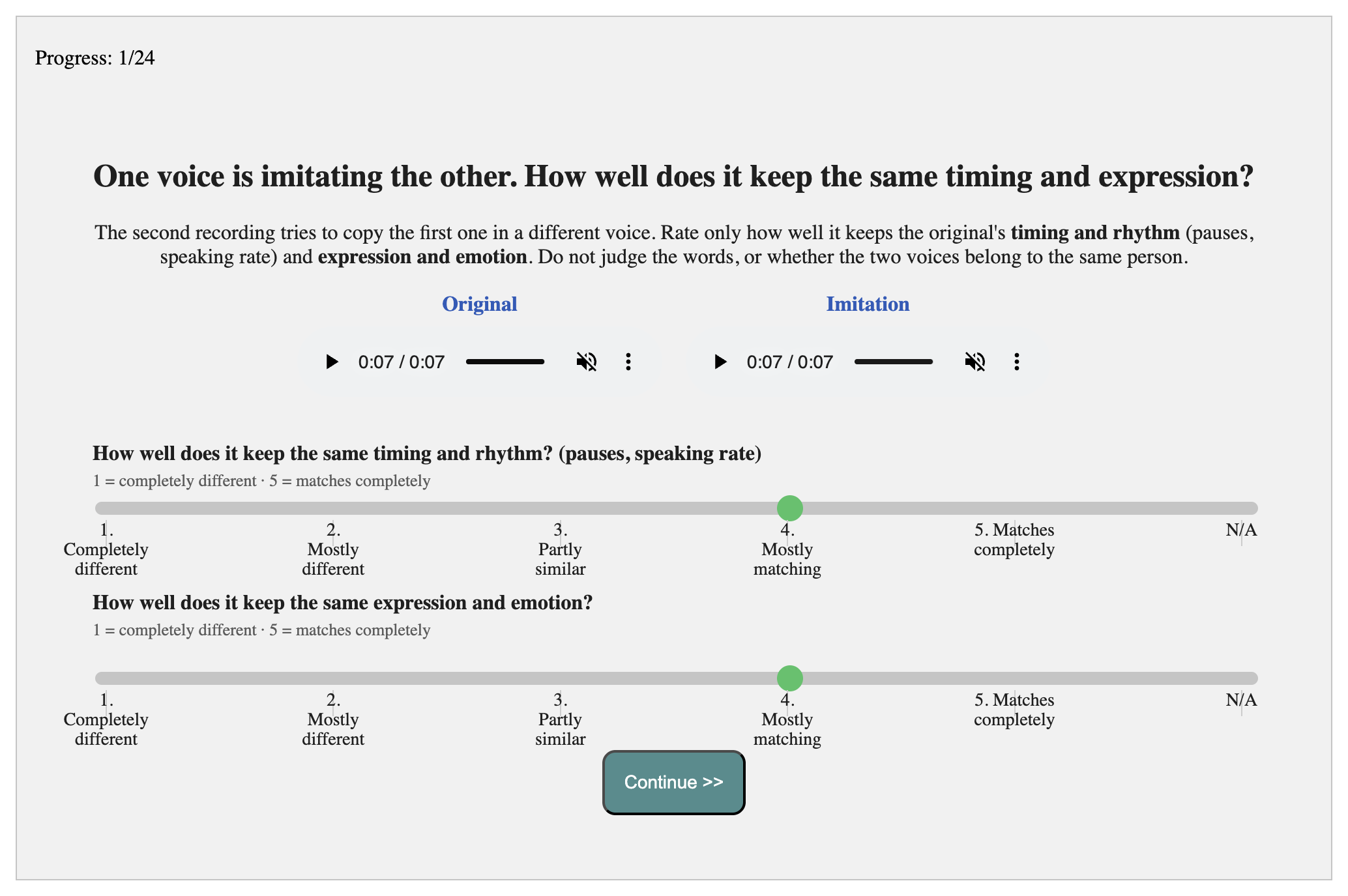}
  \caption{Source preservation (\textsc{Rhy}/\textsc{Exp}): the source is the ``Original'' and the output the ``Imitation.''}
  \label{fig:subj_vc_p}
\end{subfigure}
\caption{\textbf{English VC listening-study interface: three tasks.} VC reuses the quality (a) and
similarity (b) instruments from TTS---now $24$ items per assignment and with the \emph{target} as the
reference voice---and adds a source-preservation task (c) that compares the output against the
\emph{source}. As in TTS, no reference is shown in the quality task.}
\label{fig:subj_vc}
\end{figure}

\paragraph{Rating instruments and reference conditions.} Figures~\ref{fig:subj_instr}--\ref{fig:subj_vc}
show the English interface for every task. The \emph{naturalness and quality} task
(Fig.~\ref{fig:subj_tts}a for TTS, Fig.~\ref{fig:subj_vc}a for VC) presents a single unlabeled recording
and asks two questions---``How natural does the speech sound overall---like a real human speaker, flowing
smoothly, without choppy or halting stretches?'' (NMOS) and ``Is the speech clear and free of audio
artifacts?'' (QMOS)---each on a $1$ (Very poor) to $5$ (Excellent) scale with an N/A option. Crucially,
\emph{no reference audio is shown}: the rater hears only the system output. The \emph{similarity} task
(Fig.~\ref{fig:subj_tts}b, Fig.~\ref{fig:subj_vc}b) is the only task that reveals the target timbre: it
plays a ``Reference voice''---the TTS prompt or the VC target---next to the anonymized ``Recording'' and
asks ``How similar does the speaker in the recording sound to the reference voice?'' on a $1$ (Very
different) to $5$ (Very similar) scale, instructing raters to ``judge only the speaker's voice identity,
timbre, and accent'' and to ``ignore wording, audio quality, and loudness.'' The VC-only
\emph{source-preservation} task (Fig.~\ref{fig:subj_vc}c) frames the source as an ``Original'' and the
conversion as an ``Imitation'' and asks, on two $1$--$5$ scales, how well the imitation keeps the
original's ``timing and rhythm (pauses, speaking rate)'' and its ``expression and emotion,'' explicitly
telling raters not to judge the words or whether the two voices belong to the same person.

\paragraph{Why quality is judged without the reference.} Naturalness and quality are properties of the
generated utterance alone, so we deliberately withhold the reference clip from the quality task. We expose the
target timbre only in the similarity task and the source only in the preservation task, each with
instructions that name the single attribute to attend to. 

\paragraph{Deployed model and unified collection.} The deployed CaDiS streaming student is the
$1000$-step DMD model. For every task--language--benchmark evaluation, this model and all baselines in
the corresponding table were available before data collection and were rated together in one unified
collection using the same protocol, $200$-clip gender-balanced set, and randomized rating pages. Confidence intervals are obtained by participant-cluster bootstrap within
the same collection. Thus, every comparison is estimated directly from a common evaluation design.
Table~\ref{tab:tts_subj_zh} likewise reports the
deployed streaming student together with all Mandarin baselines from their unified collection.

For the in-domain English VC study, the streaming--offline differences are
$\Delta_{\mathrm{NMOS}}{=}{-}0.13$, $\Delta_{\mathrm{QMOS}}{=}{-}0.14$,
$\Delta_{\mathrm{SMOS}}{=}{+}0.10$, $\Delta_{\textsc{Rhy}}{=}{+}0.05$, and
$\Delta_{\textsc{Exp}}{=}{+}0.01$---all small in magnitude.

\paragraph{Paired VC comparison with X-VC.}
Table~\ref{tab:vc_xvc_paired_ci} reports paired differences between CaDiS and X-VC.
The QMOS interval lies entirely above zero, whereas the SMOS interval includes zero.
Thus, this comparison supports higher perceptual quality for CaDiS, without resolving the
speaker-similarity difference at the reported pointwise $95\%$ level.
An interval containing zero does not establish equivalence, and this cross-system comparison
does not by itself establish preservation relative to our own offline teacher.

\begin{table}[tbp]
\centering\small
\setlength{\tabcolsep}{8pt}
\caption{\textbf{Paired English VC listening-study differences: CaDiS minus X-VC}.
Positive values favor CaDiS. Intervals are pointwise $95\%$ confidence intervals.}
\label{tab:vc_xvc_paired_ci}
\begin{tabular}{lcc}
\toprule
Metric & Paired difference & $95\%$ CI \\
\midrule
QMOS & $+0.8869$ & $[+0.7143,+1.0625]$ \\
SMOS & $-0.0603$ & $[-0.2270,+0.1034]$ \\
\bottomrule
\end{tabular}
\end{table}

\begin{table}[t]
\centering\small
\setlength{\tabcolsep}{6pt}
\caption{\textbf{VC subjective listening study, Anim400K English}. Columns as in Table~\ref{tab:vc_en}. Streaming$-$offline (Ours):
$\Delta_{\text{NMOS}}{=}{-}0.16$, $\Delta_{\text{QMOS}}{=}{-}0.09$, $\Delta_{\text{SMOS}}{=}{-}0.17$,
$\Delta_{\textsc{Rhy}}{=}{+}0.12$, $\Delta_{\textsc{Exp}}{=}{+}0.16$. Objective metrics in
Table~\ref{tab:anim400k_vc}.}
\label{tab:vc_subj_anim}
\begin{tabular}{lcccccc}
\toprule
System & Stream & NMOS\,$\uparrow$ & QMOS\,$\uparrow$ & SMOS\,$\uparrow$ & \textsc{Rhy}\,$\uparrow$ & \textsc{Exp}\,$\uparrow$ \\
\midrule
\multicolumn{7}{l}{\textit{Offline references}}\\
\midrule
Vevo                              & \ding{55} & \textbf{3.83} & \textbf{3.87} & \textbf{3.76} & 4.07 & 3.74 \\
\textbf{Ours} --- offline      & \ding{55} & 3.74 & 3.82 & 3.29 & \textbf{4.13} & \textbf{3.87} \\
\midrule
\multicolumn{7}{l}{\textit{Streaming systems}}\\
\midrule
X-VC                              & \checkmark & \textbf{3.60} & 3.19 & \textbf{3.28} & 4.01 & 3.70 \\
Seed-VC                           & \checkmark & 3.19 & 3.64 & 2.92 & 3.48 & 2.97 \\
MeanVC-2                          & \checkmark & 3.28 & 3.20 & 2.59 & 3.78 & 3.36 \\
\textbf{CaDiS}   & \checkmark & 3.58 & \textbf{3.73} & 3.12 & \textbf{4.25} & \textbf{4.03} \\
\bottomrule
\end{tabular}
\end{table}
\paragraph{Anim400K (out-of-domain) VC listening study.} Table~\ref{tab:vc_subj_anim} reports the
corresponding evaluation on anime-dub voices; the three findings from the main-text English evaluation hold here too.
(i)~\textbf{Streaming matches offline on every axis}: pooling both benchmarks, all ten
streaming$-$offline differences ($2$ benchmarks $\times\,5$ axes) are small---for VC even the quality
axis is essentially unchanged ($\Delta_{\text{QMOS}}{=}{-}0.14$ English /
${-}0.09$ anime), a stronger result than TTS. (ii)~The external offline anchor \textbf{Vevo is the
speaker-similarity ceiling} (SMOS $3.76$ here, $3.67$ English), $\sim\!0.4$--$0.6$ above our systems,
while the participant-bootstrap $95\%$ CIs for its naturalness/quality gaps to ours include zero---consistent with the DMD objective carrying
no explicit speaker term. (iii)~On \textbf{source preservation our streaming student leads}
(best-in-block \textsc{Rhy}/\textsc{Exp} on both benchmarks): the live source drives content and prosody,
the flip side of its lower target similarity.

\begin{table}[t]
\centering\small
\setlength{\tabcolsep}{6pt}
\caption{\textbf{TTS subjective listening study, Seed-TTS-eval Mandarin.} Columns as in Table~\ref{tab:tts_en}; NMOS/QMOS/SMOS =
naturalness/quality/prompt-similarity. Bold = best within block. The CaDiS streaming row is the deployed DMD student. Streaming$-$offline (Ours):
$\Delta_{\text{NMOS}}{=}{-}0.21$, $\Delta_{\text{QMOS}}{=}{-}0.31$,
$\Delta_{\text{SMOS}}{=}{+}0.37$. Objective metrics in Table~\ref{tab:tts_zh}.}
\label{tab:tts_subj_zh}
\begin{tabular}{lcccc}
\toprule
System & Stream & NMOS\,$\uparrow$ & QMOS\,$\uparrow$ & SMOS\,$\uparrow$ \\
\midrule
\multicolumn{5}{l}{\textit{Offline references}}\\
\midrule
F5-TTS                                   & \ding{55} & 3.51 & 3.82 & \textbf{3.70} \\
IndexTTS-2.5                             & \ding{55} & 3.47 & 3.65 & 3.61 \\
S1-mini                                   & \ding{55} & 3.36 & 3.97 & 3.29 \\
\textbf{Ours} --- offline & \ding{55} & \textbf{3.91} & \textbf{4.07} & 3.20 \\
\midrule
\multicolumn{5}{l}{\textit{Streaming systems}}\\
\midrule
Qwen3-TTS                                & \checkmark & \textbf{3.95} & \textbf{3.99} & 3.40 \\
CosyVoice3                               & \checkmark & 3.00 & 3.76 & \textbf{3.78} \\
Faster-IndexTTS-2                        & \checkmark & 3.66 & 3.58 & 3.52 \\
\textbf{CaDiS}            & \checkmark & 3.70 & 3.76 & 3.57 \\
\bottomrule
\end{tabular}
\end{table}

\begin{table}[t]
\centering\small
\setlength{\tabcolsep}{6pt}
\caption{\textbf{VC subjective listening study, Seed-TTS-eval Mandarin} (subjective MOS, $n{=}200$
gender-balanced subset). Columns as in Table~\ref{tab:vc_en}; SMOS is vs.\ the target speaker,
\textsc{Rhy}/\textsc{Exp} are source rhythm/expression preservation. Bold = best within block.
Streaming$-$offline (Ours): $\Delta_{\text{NMOS}}{=}{-}0.12$, $\Delta_{\text{QMOS}}{=}{-}0.07$,
$\Delta_{\text{SMOS}}{=}{-}0.12$, $\Delta_{\textsc{Rhy}}{=}{+}0.06$, $\Delta_{\textsc{Exp}}{=}{+}0.22$.
Objective metrics in Table~\ref{tab:vc_zh}.}
\label{tab:vc_subj_zh}
\begin{tabular}{lcccccc}
\toprule
System & Stream & NMOS\,$\uparrow$ & QMOS\,$\uparrow$ & SMOS\,$\uparrow$ & \textsc{Rhy}\,$\uparrow$ & \textsc{Exp}\,$\uparrow$ \\
\midrule
\multicolumn{7}{l}{\textit{Offline references}}\\
\midrule
Vevo                              & \ding{55} & \textbf{3.44} & \textbf{3.78} & \textbf{3.87} & \textbf{3.90} & \textbf{3.66} \\
\textbf{Ours} --- offline      & \ding{55} & 3.43 & 3.76 & 2.97 & 3.89 & 3.57 \\
\midrule
\multicolumn{7}{l}{\textit{Streaming systems}}\\
\midrule
Seed-VC                           & \checkmark & 3.19 & 3.30 & 2.99 & 3.66 & 3.42 \\
X-VC                              & \checkmark & 3.17 & 2.63 & \textbf{3.11} & 3.79 & 3.37 \\
MeanVC-2                          & \checkmark & \textbf{3.39} & 3.43 & 2.69 & 3.78 & 3.52 \\
\textbf{CaDiS}     & \checkmark & 3.31 & \textbf{3.69} & 2.86 & \textbf{3.95} & \textbf{3.79} \\
\bottomrule
\end{tabular}
\end{table}

\paragraph{Mandarin (Seed-TTS-eval) listening study.} A global native/fluent-Mandarin Prolific audience
rated all systems together in a single collection; TTS results are in Table~\ref{tab:tts_subj_zh}
($8$ systems, no VoXtream) and VC in Table~\ref{tab:vc_subj_zh}. The streaming$\approx$offline conclusion
holds throughout: the five VC axes and the TTS naturalness axis show only small differences,
and only the TTS quality/artifact axis carries a clear (now-familiar) streaming cost
($\Delta_{\text{QMOS}}{=}{-}0.31$; naturalness $\Delta_{\text{NMOS}}{=}{-}0.21$), while on
target similarity the deployed streaming student in fact \emph{exceeds} the offline model
($\Delta_{\text{SMOS}}{=}{+}0.37$; see the prompt-similarity note below). On VC, even the quality axis is essentially unchanged
($\Delta_{\text{QMOS}}{=}{-}0.07$) and our streaming student again leads source preservation
(best-in-block \textsc{Rhy} $3.95$ and \textsc{Exp} $3.79$), while Vevo remains the speaker-similarity
ceiling (SMOS $3.87$). The one Mandarin-specific effect is a TTS prompt-similarity gap on the
\emph{offline} system, which sits last on SMOS ($3.20$) despite competitive objective WavLM-SV.

\section{Implementation and reproducibility details}
\label{app:impl}
\paragraph{Few-step sampler and noise grid.} At inference the four-step
students denoise on a uniform-in-$t$ grid with endpoints
$t\in\{1,\tfrac34,\tfrac12,\tfrac14,0\}$, snapped onto the offline teacher's
$32$-step DDIM grid ($\{32,24,16,8,0\}/32$). Because $t$ indexes the
variance-preserving shifted-cosine schedule (shift $0.5$), these equally
spaced $t$ values are \emph{not} equally spaced in noise level; the schedule
places them denser toward high SNR. The student update is stochastic rather
than deterministic (DDIM): at each endpoint $t_i\!\to\!t_{i-1}$ the student
predicts the clean latent $F_\theta(\mathbf z_{t_i},t_i)$ and \emph{re-noises}
it to $t_{i-1}$ with freshly sampled Gaussian noise, with the terminal step
$t=0$ returning the clean prediction directly. The Stage-3 DMD self-rollout
uses this same grid and update rule, so training and deployment share one
sampler.

\paragraph{Distillation chain.} The chain is a verified init-from sequence: offline bidirectional DiT
$\to$ $32$-step AR causal teacher (Stage~1) $\to$ $4$-step consistency student (Stage~2) $\to$ $4$-step
DMD student (Stage~3). For DMD, the real (frozen) score is the offline teacher of the corresponding task
(the bidirectional generator for TTS, the offline content-addition converter for VC), and the DMD
gradient is applied directly to the selected generated chunks, while cached history is pinned as
detached conditioning. The consistency
student initializes the DMD student (asymmetric initialization).

\paragraph{Sequence lengths and self-rollout.} Target lengths are capped per task and stage (at $40\hz$, $40$ frames $=1$\,s). For \emph{TTS}, the AR teacher and the consistency student are trained on targets up to $30$\,s ($1200$ frames), while the Stage-3 DMD student is distilled with self-rollout on $15$\,s clips ($600$ frames); every valid chunk receives a direct DMD gradient. For \emph{VC}, the teacher and the consistency student are trained on targets up to $15$\,s ($600$ frames), and the DMD student likewise rolls out over $15$\,s, but only the $24$ most recent $400$\,ms blocks ($\sim$$9.6$\,s) receive direct DMD gradients. In both tasks, each generated chunk is detached before being committed to the causal cache, so these direct gradients do not propagate through earlier chunks as recurrent history. Thus, ``all chunks'' for TTS and ``the last $24$ chunks'' for VC describe the support of the loss, not cross-chunk back-propagation.

\paragraph{Training recipe.} The offline bidirectional teacher (DiT backbone,
$0.8$B parameters) is trained to $500$k steps. From it we train the $32$-step AR causal teacher (Stage~1)
for $50$k steps on $32$ GPUs, distill the $4$-step consistency student (Stage~2) for $30$k steps, and reach
the $4$-step DMD student (Stage~3), initialized from the consistency student, in only
$1000$ steps; this $1000$-step Stage-3 budget is the same for both the TTS and VC students. All three causal stages share the 400 ms block configuration above; Stages 2 and 3 use four NFEs, while Stage 1 uses 32. Table~\ref{tab:training_hparams} gives the task- and stage-specific optimization settings.
We use AdamW with weight decay $0.01$ throughout and the default
$(\beta_1,\beta_2){=}(0.9,0.999)$, except for the DMD critic, which uses
$(\beta_1,\beta_2){=}(0,0.999)$. For TTS, each microbatch concatenates examples from four data
sources; the parenthesized values show their contributions. At Stage~3, four single-example
microbatches are accumulated into one local batch. The reported global batch is the number of
examples contributing to one optimizer update after accumulation.

\begin{table}[h]
\centering\small
\setlength{\tabcolsep}{4pt}
\caption{Training optimization settings. For DMD, the two learning rates are for the
generator and critic, respectively. ``Accum.'' is the number of microbatches per optimizer update.}
\label{tab:training_hparams}
\begin{tabular}{llccccc}
\toprule
Task & Stage & Learning rate & Microbatch/GPU & Accum. & GPUs & Global batch \\
\midrule
TTS & Offline teacher & $1\!\times\!10^{-4}$ & $36$ $(8{+}8{+}12{+}8)$ & 1 & -- & -- \\
TTS & AR teacher      & $3\!\times\!10^{-5}$ & $18$ $(4{+}4{+}6{+}4)$  & 2 & 32 & 1152 \\
TTS & Consistency     & $2\!\times\!10^{-6}$ & $18$ $(4{+}4{+}6{+}4)$  & 2 & 32 & 1152 \\
TTS & DMD             & $2\!\times\!10^{-6}$ / $4\!\times\!10^{-7}$ & 1 & 4 & 64 & 256 \\
\midrule
VC  & Offline teacher & $1\!\times\!10^{-4}$ & 8 & 1 & -- & -- \\
VC  & AR teacher      & $1\!\times\!10^{-4}$ & 8 & 1 & 32 & 256 \\
VC  & Consistency     & $4\!\times\!10^{-6}$ & 4 & 1 & 32 & 128 \\
VC  & DMD             & $2\!\times\!10^{-6}$ / $4\!\times\!10^{-7}$ & 1 & 1 & 64 & 64 \\
\bottomrule
\end{tabular}
\end{table}

\paragraph{Latency measurement.} All FPL/RTF numbers are measured on an idle A100-80GB, with warm-up
iterations discarded and the median reported ($n{\ge}44$). We report FPL as algorithmic buffering
(chunk $+$ each system's own look-ahead) plus measured compute-to-first-packet; RTF is wall time divided
by audio duration. CPU latency numbers from prior work are not directly comparable to these GPU numbers,
and we always name the hardware.
\paragraph{TTS input availability and timing scope.}
The efficiency of all TTS systems are computed with the complete target sentence available before inference, including incremental-capable baselines such as VoXtream2. Timing includes the text front end, first-chunk generation, and waveform decoding;
CaDiS additionally includes duration estimation. Model loading and warm-up are excluded.
The comparison concerns full-text-ready synthesis, not incremental-text capabilities.

\begin{table}[tbp]
\centering
\footnotesize
\setlength{\tabcolsep}{4pt}
\renewcommand{\arraystretch}{1.15}
\caption{\textbf{Streaming engines and chunk/context configurations}.
All chunk, window, overlap, and look-ahead durations are expressed in milliseconds of audio.}
\label{tab:streaming_engine_settings}
\begin{tabular}{@{}p{0.22\linewidth}p{0.30\linewidth}p{0.42\linewidth}@{}}
\toprule
System & Engine / acceleration & Chunk / context (ms of audio) \\
\midrule
Qwen3-TTS
& \texttt{faster-qwen3-tts}; CUDA graphs
& 640 \\

VoXtream2
& Native CUDA graphs; Mimi
& 80 \\

CosyVoice3
& Official vLLM; FP32 flow/vocoder
& 1000 chunk; hop $1000\to2000\to4000$; 120 look-ahead \\

Faster-IndexTTS-2
& TensorRT / TensorRT-LLM
& $\approx2000$ window; $\approx93$ overlap \\

Seed-VC
& PyTorch
& 250 block + 50 crossfade + 20 right context \\

X-VC
& PyTorch
& 120 current + 20 smoothing + 100 future \\

MeanVC-2
& PyTorch
& 120 current + 40 future \\

CaDiS--TTS
& PyTorch
& 400 \\

CaDiS--VC
& PyTorch
& 400; no future look-ahead \\
\bottomrule
\end{tabular}
\end{table}

\paragraph{Sampling rates and parameter counts.}
Training dataset consists of audio mostly at 48kHz, for audio with a sample rate of lower than 48kHz, we resampled to $48$\,kHz before model input. Both CaDiS pipelines produce $48$\,kHz audio, and their RTF measurements include waveform decoding at 48kHz.
Baseline timing retains each system's native output sampling rate.

The complete TTS system contains approximately $1.5$B parameters: an approximately $0.8$B DiT,
a $564$M mT5 text encoder, and a $128$M DAC-VAE. The VC system contains approximately $0.65$B
parameters: a $421$M DiT, a $94$M ContentVec encoder, and the same $128$M DAC-VAE.
These totals include frozen components.

\paragraph{Chunked waveform decoding.}
The DAC-VAE decoder is not sample-causal. We reuse the offline DAC-VAE and implement streaming
waveform generation by decoding latent chunks and crossfading adjacent decoded segments.
\paragraph{Scoring.} All WER/CER/SIM numbers flow through the Seed-TTS-eval scripts (Whisper-large-v3;
Paraformer-zh; WavLM-large-SV) plus Resemblyzer, SimAM-ResNet100/VoxBlink2, and UTMOS. Long-form drift
uses an overlap-window ASR to avoid a non-overlapping-chunk artifact, and windowed Resemblyzer for the
early$\to$late similarity slope.

\end{document}